\documentclass[preprint2]{aastex}
\usepackage{pdflscape}
\usepackage{amssymb}
\usepackage{graphicx}
\usepackage{natbib}
\usepackage{lscape}
\usepackage{rotating}
\usepackage{txfonts}

\newcommand{\kms}{km\ s$^{-1}$}

\newcommand{\msol}{$M_{\odot}$}

\newcommand{\xmm}{{\sc XMM}\emph{-Newton}}

\newcommand{\ch}{\emph{Chandra}}
\newcommand{\lxlb}{$L_{\rm X}/L_{\rm BOL}$}
\newcommand{\loglxlb}{$\log[L_{\rm X}/L_{\rm BOL}]$}

\shorttitle{X-ray from magnetic stars}
\shortauthors{Naz\'e et al.}

\begin{document}


\title{X-ray emission from magnetic massive stars\thanks{Based on data collected with \xmm, and \ch.}}

\author{Ya\"el Naz\'e\altaffilmark{*}}
\affil{GAPHE, D\'epartement AGO, Universit\'e de Li\`ege, All\'ee du 6 Ao\^ut 17, Bat. B5C, B4000-Li\`ege, Belgium}
\email{naze@astro.ulg.ac.be}

\author{V\'eronique Petit}
\affil{Dept. of Physics \& Space Sciences, Florida Institute of Technology, Melbourne, FL, 32901, USA}

\author{Melanie Rinbrand}
\affil{Department of Physics \& Astronomy, University of Delaware, Bartol Research Institute, Newark, DE 19716, USA}

\author{David Cohen}
\affil{Department of Physics \& Astronomy, Swarthmore College, Swarthmore, PA 19081, USA}

\author{Stan Owocki}
\affil{Department of Physics \& Astronomy, University of Delaware, Bartol Research Institute, Newark, DE 19716, USA}

\author{Asif ud-Doula}
\affil{Penn State Worthington Scranton, Dunmore, PA 18512, USA}

\and
\author{Gregg A. Wade}
\affil{Department of Physics, Royal Military College of Canada, PO Box 17000, Station Forces, Kingston, ON K7K 4B4, Canada}

\altaffiltext{*}{Research Associate FRS-FNRS}


\begin{abstract}
Magnetically confined winds of early-type stars are expected to be sources of bright and hard X-rays. To clarify the systematics of the observed X-ray properties, we have analyzed a large series of \ch\ and \xmm\ observations, corresponding to all available exposures of known massive magnetic stars (over 100 exposures covering $\sim$60\% of stars compiled in the catalog of Petit et al. 2013). We show that the X-ray luminosity is strongly correlated with the stellar wind mass-loss-rate, with a power-law form that is slightly steeper than linear for the majority of the less luminous, lower-${\dot M}$ B stars and flattens for the more luminous, higher-${\dot M}$ O stars. As the winds are radiatively driven, these scalings can be equivalently written as relations with the bolometric luminosity. The observed X-ray luminosities, and their trend with mass-loss rates, are well reproduced by new MHD models, although a few overluminous stars (mostly rapidly rotating objects) exist. No relation is found between other X-ray properties (plasma temperature, absorption) and stellar or magnetic parameters, contrary to expectations (e.g. higher temperature for stronger mass-loss rate). This suggests that the main driver for the plasma properties is different from the main determinant of the X-ray luminosity. Finally, variations of the X-ray hardnesses and luminosities, in phase with the stellar rotation period, are detected for some objects and they suggest some temperature stratification to exist in massive stars' magnetospheres. 
\end{abstract}


\section{Introduction}
As they lack deep outer convective envelopes, dynamos analogous to the Sun's are not expected to operate in early-type stars (A, B, O). Magnetic fields were however found in a few percent of the population of main sequence A and late B-type stars in the Galaxy \citep{wol68,pow07} and, more recently, in a similar proportion of early B and O stars \citep[and references therein]{hub11,wad13}. While they are most probably fossil fields, their detailed origin remains elusive (primordial field, early dynamo, binary mergers, \citealt{fer09,bra13,lan13}). The detected magnetic fields share similar properties: they are generally strong (a few kG), organized on large scales (i.e., with important dipole component), and globally stable on timescales of at least years. 

Such strong, organized magnetic fields are able to channel the stellar wind flows towards the magnetic equator, giving rise to regions of magnetically confined wind (MCW) and creating a stellar magnetosphere \citep{sho90,bab97,udd02}. The presence of these dense confined winds leads to additional emissions or absorptions throughout the electromagnetic spectrum. In the high-energy range, X-rays should arise from the collision between the high-velocity wind flows channeled from both hemispheres along the magnetic field lines \citep[e.g. ][]{bab97}.

Indeed the observed properties of $\theta^1$\,Ori\,C (O7Vfp, \citealt{ku82,shu00,gag05}) agree well with theoretical expectations \citep{bab97theta,gag05}. The X-ray emission is overluminous by one dex compared to similar non-magnetic stars and is dominated by a thermal component at $\sim$3\,keV, compared to 0.2--0.6\,keV in `normal' O stars.  Furthermore, the X-ray lines are narrow and on average unshifted while the line ratios indicate a formation region close to the photosphere, as expected for slow-moving material trapped in a magnetosphere. Simultaneous X-ray and optical variations further underline the link between MCWs and X-rays \citep{gag97,gag05}.  While some details of the observations are not yet reproduced by models\footnote{The emitting plasma is located too close to the photosphere; unexplained variations of the X-ray absorption are observed as the viewing angle of the magnetosphere changes; the observed X-ray line widths are slightly too large; small shifts of X-ray lines are observed throughout the rotation cycle, but are not predicted \citealt{gag05}}, the overall agreement is still quite satisfactory. $\theta^1$\,Ori\,C thus appears as a prototype for understanding magnetospheres of slowly-rotating massive stars.

For cooler and more rapidly rotating objects, $\sigma$\,Ori\,E (B2Vp) appears as another well-studied landmark. Its emission is also hard and luminous in the X-ray range \citep{san04,ski08}, although no details on its X-ray lines are yet available. The properties of $\sigma$\,Ori\,E are roughly reproduced by models \citep{tow07} and even the predicted magnetic braking has been detected observationally for this object \citep{tow10}. 

These two prototypes are not the sole magnetic objects observed in the X-ray range. However, the other objects conform less well to the theoretical expectations. The magnetic Of?p stars display order-of-magnitude X-ray overluminosity, some narrow X-ray lines, and correlated X-ray/optical changes but their high-energy emission is dominated by soft X-rays rather than hard ones \citep{naz04,naz07,naz08,naz10,naz12}.  $\tau$\,Sco (B0.2V) displays an overluminosity, a relatively hot (1.7\,keV) component, narrow and unshifted X-ray lines formed close to the star \citep{mew03,coh03}. However, the relatively complex magnetic topology of the star \citep{don06} was expected to generate strong variability of the X-ray emission with the rotational period, which was not observed \citep{ign10}. Finally, \citet{osk11} compared the X-ray properties of a sample of 11 magnetic B stars, including $\sigma$\,Ori\,E (see above), $\beta$\,Cep \citep{fav09}, LP\,Ori and NU\,Ori \citep{ste05}, while \citet{ign13} analyzed new X-ray observations of two $\tau$\,Sco analogs. The situation again appears quite varied: some objects displayed hard X-ray emission, while others rather emit soft X-rays; overluminosity seemed to be the exception rather than the rule. No obvious correlation was found by \citet{osk11} between the high-energy properties and bolometric luminosity, magnetic field strength, rotation period, or pulsation period (when existing). Therefore, the origin of the discrepancy between observations and models remains unknown. 

The situation thus appears less satisfactory than the few iconic magnetic objects would at first suggest. In order to clarify the situation, the detailed correlation between magnetic/stellar properties and the X-ray characteristics should be assessed systematically for a larger sample of stars, notably searching for potential differences in X-ray observables related to their magnetospheric structure. To do so, we examine the overall sample of magnetic massive stars whose stellar and magnetic parameters are generally well known \citep{pet13}. Section 2 presents the X-ray observations used in this study and the analysis method, Sections 3 and 4 describe the observational results and their interpretation while Section 5 summarizes our findings and concludes this paper.

\section{Observations}
Several X-ray observatories have flown since the 1970s. However, their instruments had various capabilities, not always comparable. In this study, we aim to maximize the number of detections while ensuring the highest possible homogeneity in the analysis (i.e. similar spectral resolution and energy band). We searched {\em Swift} and {\em Suzaku} archives for observations of the magnetic stars of \citet{pet13} but, with the exception of $\tau$\,Sco \citep{ign10}, no source was detected by these facilities. We also found {\em ASCA} detections for five of our targets, but these objects lie in clusters, where the coarse PSF of {\em ASCA} makes it difficult to extract uncontaminated data. We thus focused our work on CCD spectra in the 0.5--10.0\,keV band, which allows us to detect hard emissions. Our analysis is based on \xmm-EPIC and \ch-ACIS data, with a mix of targeted programs (notably our own programmes, PI Naz\'e and PI Petit) and serendipitous archival observations. There are more than a hundred exposures available for 40 targets. Therefore X-ray observations are available for 63\% of the Petit et al. catalog. Table \ref{listoftargeta} provides the stellar/magnetic properties of the sample, from \citet{pet13}, whereas Table \ref{listoftargetb} provide the detailed information on the X-ray observations. 

\begin{figure}
\includegraphics[width=8.5cm]{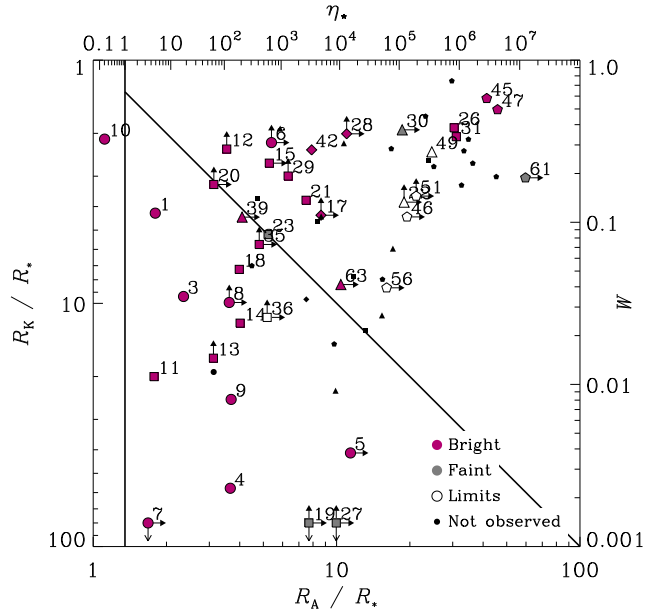}
\caption{Location of the targets in the magnetic confinement-rotation diagram \citep[see Fig. 3 of ][for the identification of all individual stars]{pet13}. The confinement parameter $\eta_*=B_p^2 R_*^2/ (4 \dot M v_{\infty})$ and the Alfven radius $R_A(R_*)\sim0.3+(\eta_*+0.25)^{1/4}$ are given as top/bottom abscissa, respectively. The right and left ordinates show the ratio of rotation speed to orbital speed $W$ and the Kepler corotation radius $R_K=W^{-2/3} R_*$, respectively \citep[see e.g.][for more discussion about these parameters]{pet13}. Highly confined winds of rapidly rotating stars therefore appear at the top right of the diagram. The symbol shapes represent spectral types (circles: O-type stars, squares: B-type stars with $T_{eff}>22$\,kK, triangles: B-type stars with 19$<T_{eff}<22$\,kK, pentagons:B-type stars with $T_{eff}<19$\,kK, diamonds: Herbig Be stars). Darker symbols correspond to objects with brighter X-ray emission (see Sect. 2: ``bright'' objects are those studied spectroscopically). }
\label{ipodfig}
\end{figure}

It is of interest to examine the location of our sample in the magnetic confinement-rotation diagram describing the dynamical structure of magnetospheres (Fig. \ref{ipodfig}, see also \citealt{pet13} for further details). Although we do not have observations of the entire catalog of \citet{pet13}, our targets are well distributed: we are thus not sampling a particular subpopulation amongst the known magnetic massive stars, and our conclusions on the magnetic OB stars should therefore have a general character. 

\xmm\ data were reduced with SAS v13.0.0 using calibration files available in June 2013 and following the recommendations of the \xmm\ team\footnote{SAS threads, see \\ http://xmm.esac.esa.int/sas/current/documentation/threads/ }. Data were filtered for keeping only best-quality data ($PATTERN$ of 0--12 for MOS and 0--4 for pn) and discarding background flares affecting the observations. A source detection was performed on each EPIC dataset using the task {\it edetect\_chain} on the 0.4--10.0\,keV energy band and for a likelihood of 10. This task searches for sources by using a sliding box and determines the final source parameters from point-spread-function (PSF) fitting: the final count rates correspond to equivalent on-axis, full PSF count rates. When sources displayed high count rates, the possibility of pile-up was assessed using the pattern distribution (task {\it epatplot}): datasets with non-negligible pile-up  were discarded and do not appear in Table \ref{listoftargetb}. For the remaining observations, we then extracted EPIC spectra using the task {\it especget} for circular regions centered on the best-fit positions of the sources and regions as close as possible to the targets for the backgrounds. The background positions as well as the extraction radii were adapted taking into account the crowding near the source as well as the off-axis PSF degradation. EPIC spectra were grouped, using {\it specgroup}, to obtain an oversampling factor of five and to ensure that a minimum signal-to-noise ratio of three (i.e. a minimum of 10 counts) was reached in each spectral bin of the background-corrected spectra. Note that for $\sigma$\,Ori\,E, only the events outside the flare were considered, as this flare is probably due to a low-mass companion \citep{san04,bou09}.

The \textit{Chandra} ACIS observations were reprocessed following the standard reduction procedure with \textsc{ciao} version 4.5\footnote{CIAO threads see http://cxc.harvard.edu/ciao/threads/.}. The procedure is similar to the one described for EPIC observations. Source where searched with the \textit{celldetect} tool and the count rates where determined with \textit{aprates}. The ACIS spectra and responses where extracted using the standard \textit{specextract} with regions centred on the source with background regions as close as possible to the target. The spectra were grouped in the same fashion as the EPIC spectra. The presence of pile-up was estimated from the count rate per frame, and exposures with non-negligible pile-up were discarded.

Of the 40 magnetic massive stars with X-ray observations, six B stars (ALS3694, HD\,55522, HD\,36485, HD\,306795, HD\,37058, HD\,156424) remain undetected while five other ones (Tr16-13, HD\,163472, ALS15956, HD\,37017, HD\,175362) have only very faint detections and the last object, ALS8988, a questionable detection\footnote{The X-ray detection of ALS8988 was first reported by \citet{tow03} but a longer \ch\ exposure resolved the emission into two sources, each separated by 2'' from the B star position, casting doubt on the detection of X-rays from ALS8988 \citep{wan08}. We thus discard this source from our sample and do not discuss it further.}. For \xmm, the equivalent on-axis count rates associated with faint detections as well as the 90\% detection limits throughout the field-of-view were automatically calculated during the source detection process. For the \ch\ observations, the number of counts, count rates, and their associated 1$\sigma$ errors at the position of these targets were estimated from aperture photometry by the task {\it aprates}, with a Bayesian estimation of the background. The one-sided, 90\% upper limits are taken as 1.28$\sigma$ in case of non-detection. Corrections for incomplete PSF and effective area from the \ch\ manual were then applied to obtain the final estimate of the upper limits\footnote{A factor of two correction was also applied to HD\,36485, which was observed with ACIS-S+HETG (resulting from the redirection of 50\% of the flux into towards the gratings)}. The \ch\ and \xmm\ count rates were then converted into fluxes (corrected for ISM absorption) using WEBPIMMs. To this aim, we used models combining the individual interstellar absorption (without additional absorption) and one thermal component. For the latter, several temperatures between 0.3 and 2.0\,keV, as found suitable for most other B stars (see below), were tried and they yielded comparable results. Table \ref{faint} provides the derived X-ray luminosities and \lxlb\ ratios.

A total of 28 magnetic stars have at least one X-ray spectrum usable for spectral modelling (i.e. 44\% of the Petit et al. catalog). These spectra were fitted within Xspec v12.7.0 with the aim of using homogeneous fitting procedures to get homogeneous and comparable results. We fitted the spectra using absorbed optically-thin thermal plasma models, i.e. $wabs \times phabs \times \sum apec$, with solar abundances \citep{and89}. The first absorption component represents the interstellar column, which was fixed to $5.8\times10^{21}\times E(B-V)$\,cm$^{-2}$ \citep[see Table \ref{listoftargeta}]{boh78}. The second absorption allows for possible additional (local) absorption, e.g. due to the stellar wind (confined or not). Regarding the emission components, two methods were used. The first uses one or two optically-thin thermal models with free temperatures. Two thermal components were only used if a single component did not provide a satisfactory fit. In this case, input temperatures of 0.45 and 1.0\,keV were used as first guesses. The second method considers a given set of four absorbed optically-thin thermal plasma models, this time with fixed temperatures of 0.2, 0.6, 1.0, and 4.0\,keV. This set of temperatures was chosen to minimize erratic results and to ensure a good representation of the X-ray emissions of the magnetic stars. Tables \ref{4tfits}, \ref{2tfits}, and \ref{observable} provide the parameters of the best fits. We also provide the observed fluxes in 3 energy bands and their associated fluxes corrected for the interstellar absorption, with $1\sigma$ error bars\footnote{These errors were calculated using Xspec {\it error} command for spectral parameters or {\it flux err} command for observed fluxes. When the error bars were asymmetric, the largest value is given here. As is usual in Xspec, these errors do not adequately take into account the interactions between spectral parameters and errors on dereddened fluxes cannot be formally calculated (as some information is missing, e.g. the error coming from the good/bad choice of model). Note that unrealistically large errors may sometimes be derived, especially when the additional absorption is close to zero. }. It must be noted that the results described in the next sections are consistent when using either fitting method.

For the brightest objects in the sample, the reduced $\chi^2$ of the best-fit are sometimes larger than 2, hence are formally not acceptable. We however kept these results notably because, even in these cases, the fitting was good in the 0.5--10.0\,keV energy band where the flux was estimated. Most of the problems in such cases probably results from nonsolar abundances and/or unrealistically small error bars (that do not account for the calibration systematics). In some cases, we fixed one or several spectral parameters. This typically happened when: (1) the additional absorption was very low ($<10^{19}$\,cm$^{-2}$) and its associated error was unrealistically large ($\sim10^{23}$\,cm$^{-2}$) - the absorption was then fixed to zero, and (2) several spectra of the same target were available (see Sect. 4.3) - we identified the best-fit parameters (absorption and/or temperature(s) and/or normalizations, depending on the source's behaviour) which were not significantly varying throughout exposures and we used them as fixed parameters for a second fit of the individual exposures. This allows us to clarify the observed trends by avoiding erratic results (especially when individual spectra display a low number of recorded counts) and removing the uncertainty in temperature coming from its interplay with absorption.

Once X-ray spectra have been fitted, the next step is to evaluate how the X-ray emission of magnetic stars relates to the MCW phenomenon, through the comparison of specific parameters. 
Three main observables are available for each spectral fitting: the level of X-ray emission, its spectral shape, and its absorption. We examine each one in turn in the following sections, and then determine the variability properties of the targets, when possible. 

\section{The X-ray luminosity of magnetic massive stars}
\subsection{Observational results}

\begin{figure}
\includegraphics[width=8.5cm,bb=30 150 420 700, clip]{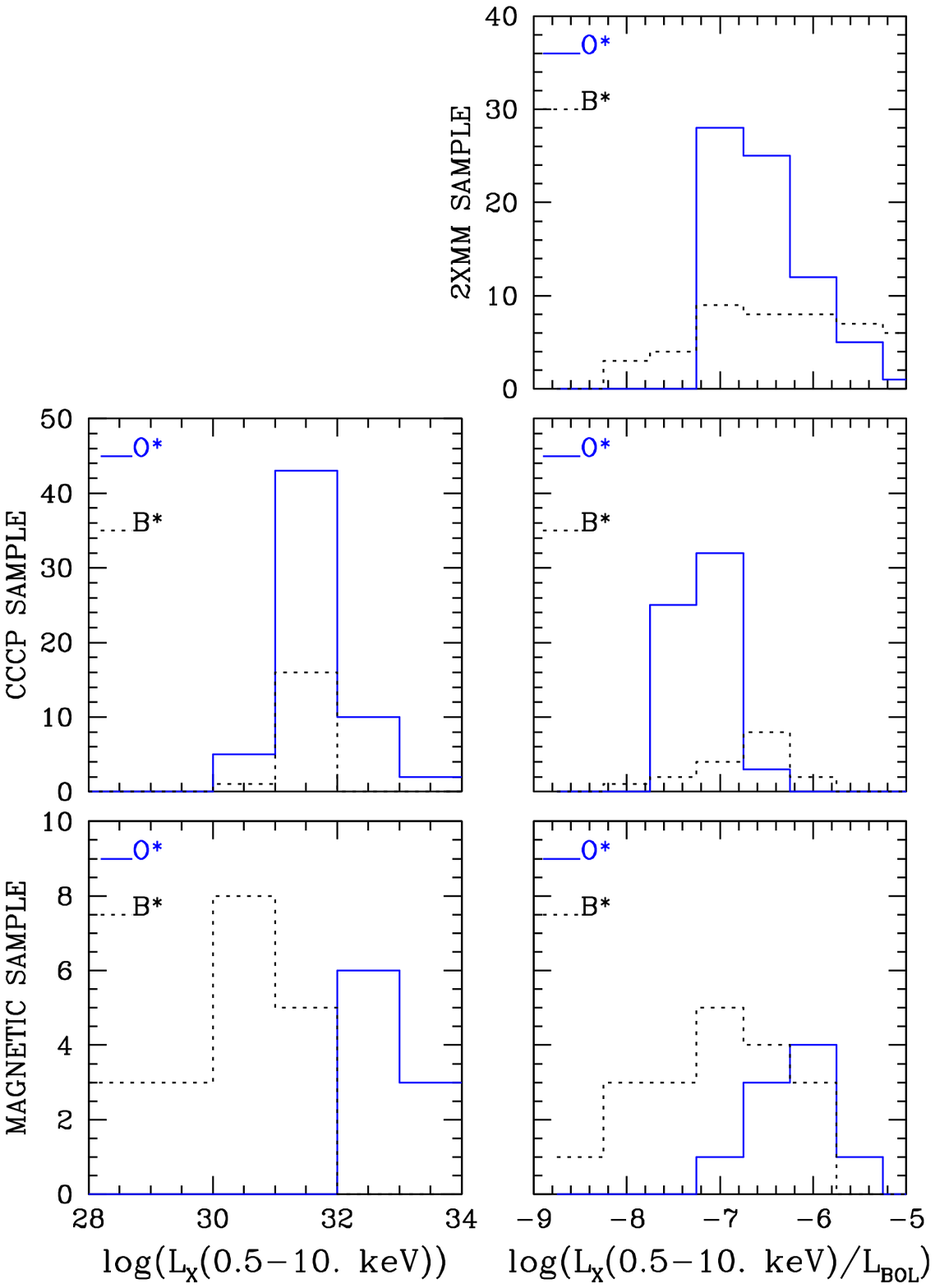}
\caption{X-ray luminosities (corrected for ISM absorption) and \lxlb\ ratios of massive stars in 2XMM \citep[top panel]{naz09} and CCCP \citep[middle panels]{naz11} compared to values for stars in this paper (4T fits, bottom panels). The solid blue lines refer to O stars, and the black dotted ones to B stars. Note that the shift in average \loglxlb\ between CCCP and 2XMM comes from the different analysis choices \citep[see][for a discussion of this problem]{naz11}.}
\label{histo}
\end{figure}

In our sample, we observed a large range of X-ray luminosities and \loglxlb\ ratios. Fig. \ref{histo} compares the observed values for our sample of magnetic massive stars with two large samples of OB stars: Chandra Carina Complex Project (CCCP, \citealt{naz11}) and the 2XMM survey \citep{naz09}\footnote{Note that these samples are not fully representative of non-magnetic single stars, as they are contaminated by a few X-ray bright colliding wind binaries and a few magnetic objects (which are also included in our sample).}. There are some obvious differences.

First, the X-ray observations of a particular region (e.g. Carina, but see also NGC6231, \citealt{san06}) have rather high detection limits, leading to a high cut-off in the luminosities: while this generally has little impact on O star detectability, many B stars with faint X-ray emission are thus missed. In our sample, some deeper exposures are available, notably because of observations acquired by the authors, and this leads to the detection of much fainter X-ray emission (Fig. \ref{histo}).  Second, for ``normal'' single, non-magnetic O stars, embedded wind shocks lead to \lxlb\ of about $10^{-7}$ \citep[Fig. \ref{histo},][and references therein]{ber97,naz09,naz11}  while larger values are observed for the magnetic stars.

\begin{figure}
\centering
\includegraphics[width=8cm]{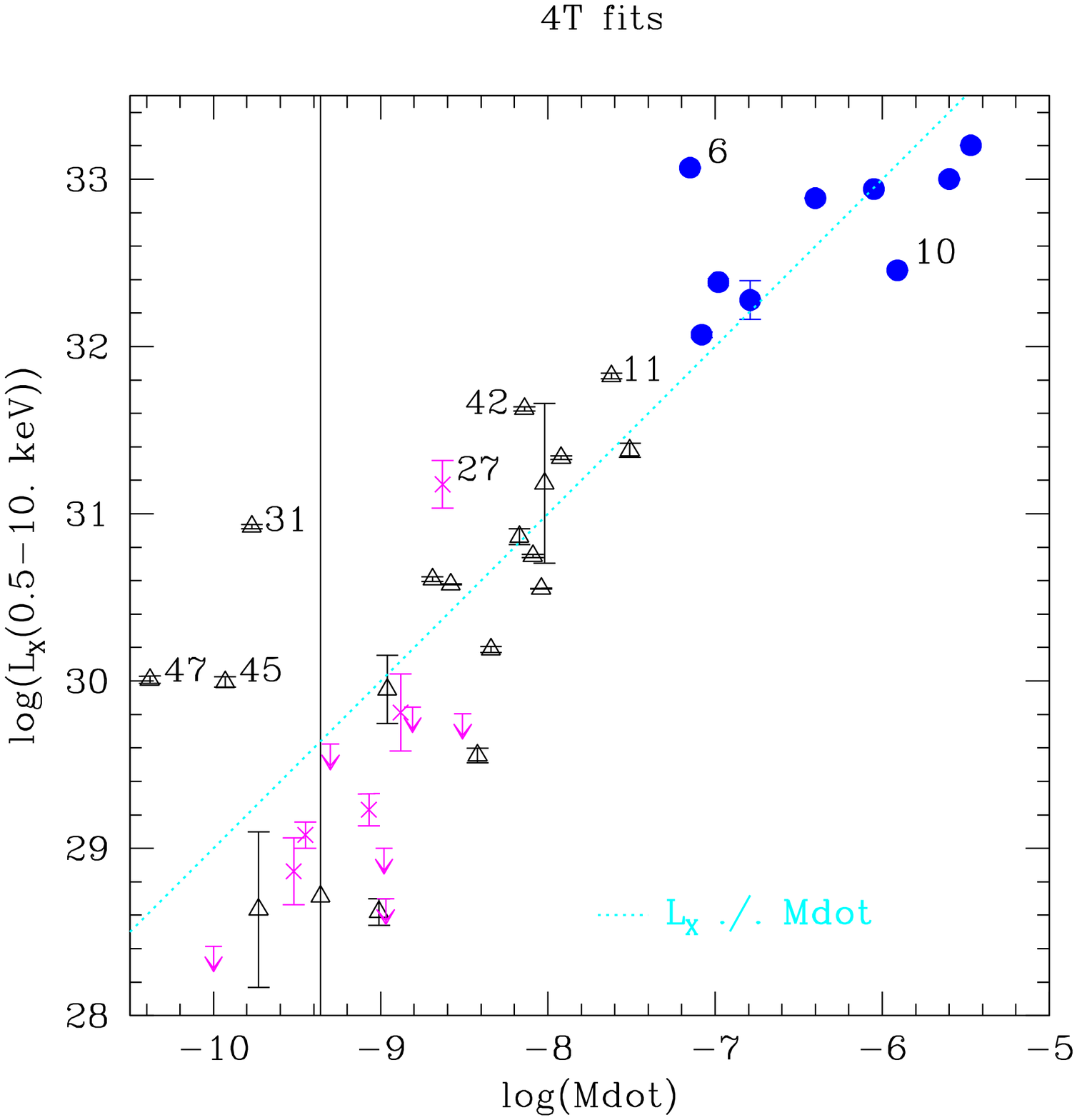}
\includegraphics[width=8cm]{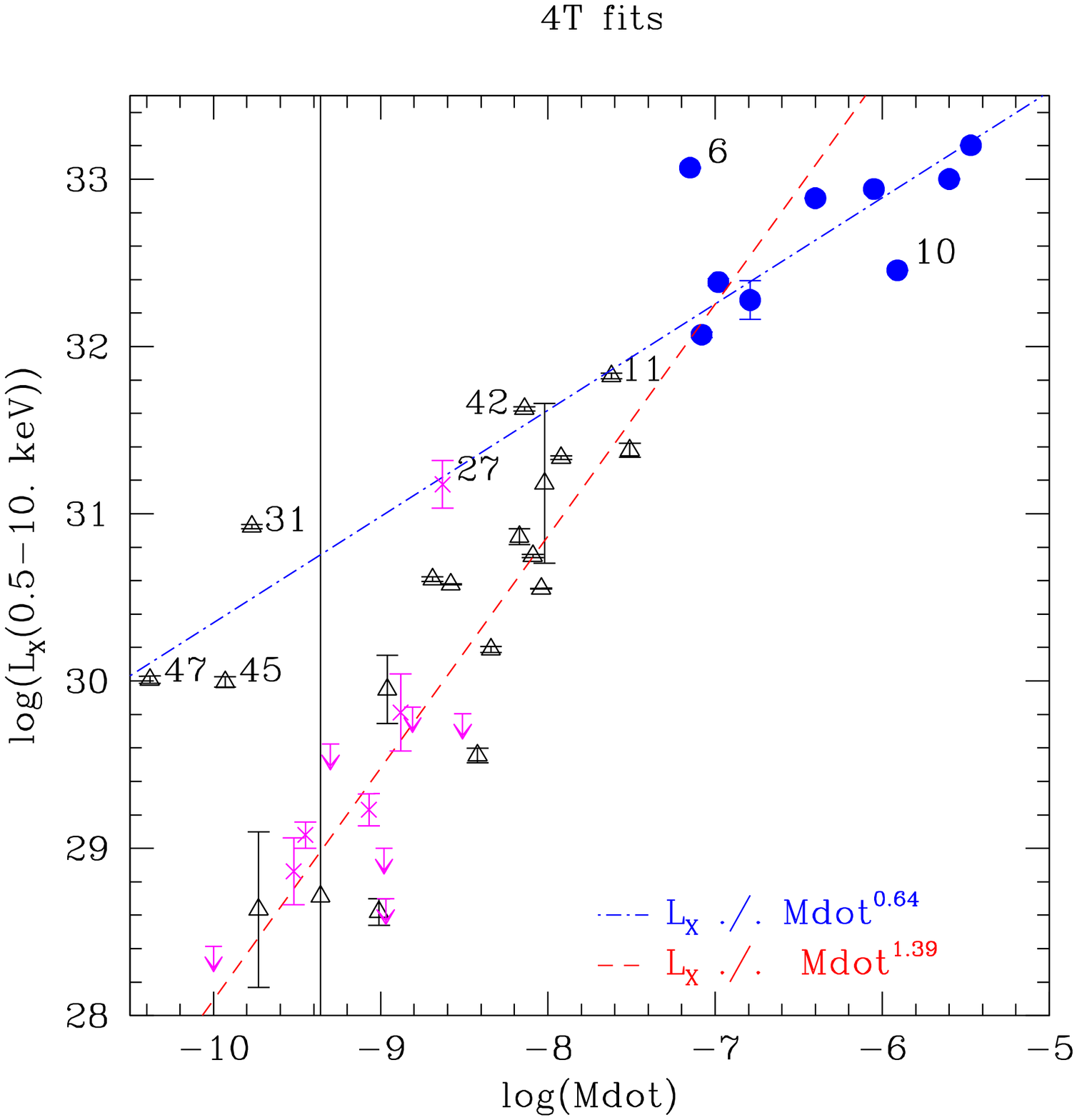}
\caption{X-ray luminosity (corrected for ISM absorption, from 4T fits) as a function of mass-loss rate. Filled blue dots correspond to O stars, black empty triangles to B stars, magenta crosses and downward-pointing arrows to faint detections and upper limits on the X-ray luminosity, respectively. The labelled line on the top panel illustrates a $L_{\rm X} \propto \dot M$ relation, whereas we show on the bottom panel the best-fit relations (see text and Table \ref{relations}). Stars of particular interest are labeled according to their identification number in Table \ref{listoftargeta}.}
\label{mdot}
\end{figure}

\begin{figure}
\centering
\includegraphics[width=8cm]{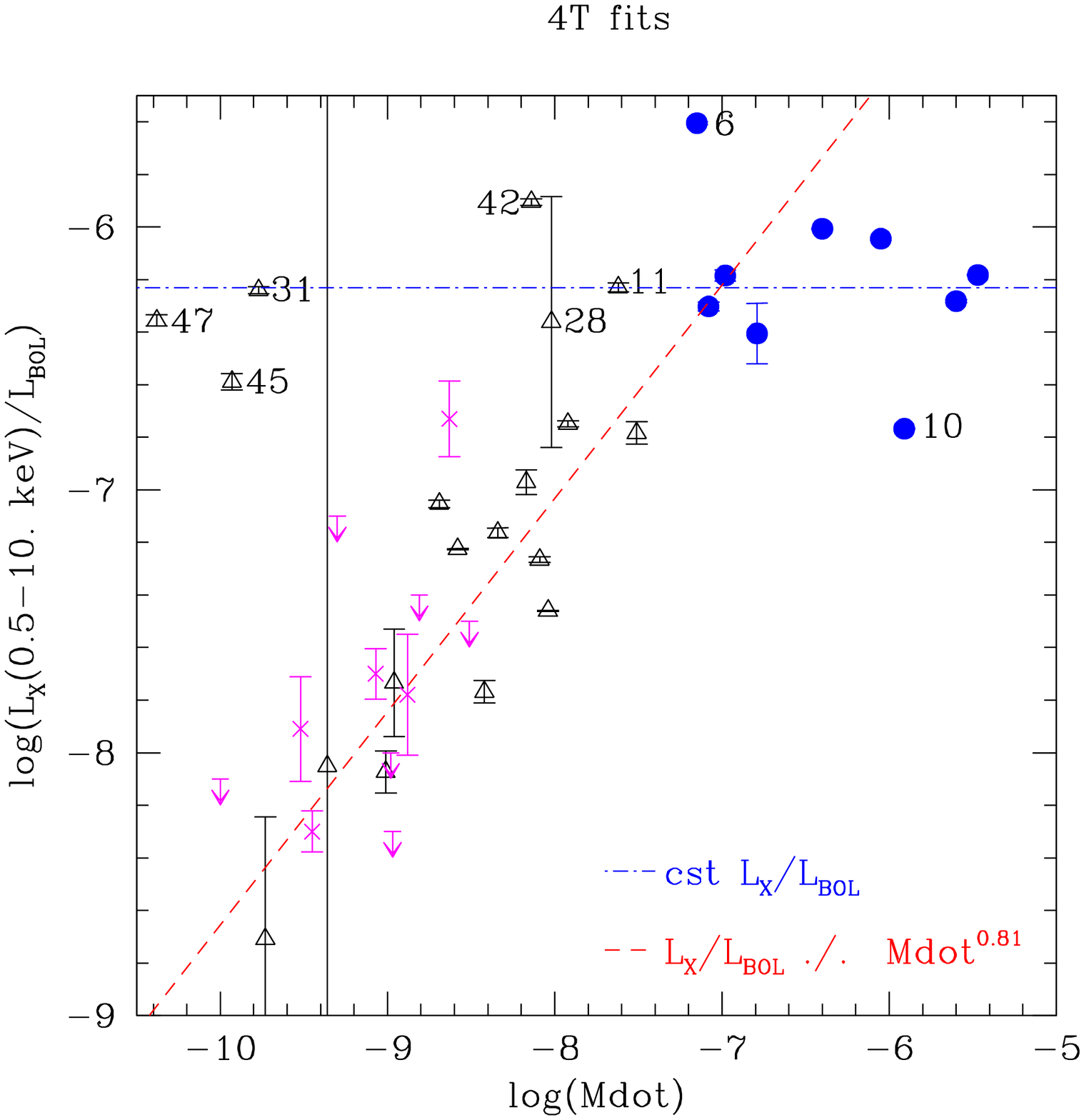}
\includegraphics[width=8cm]{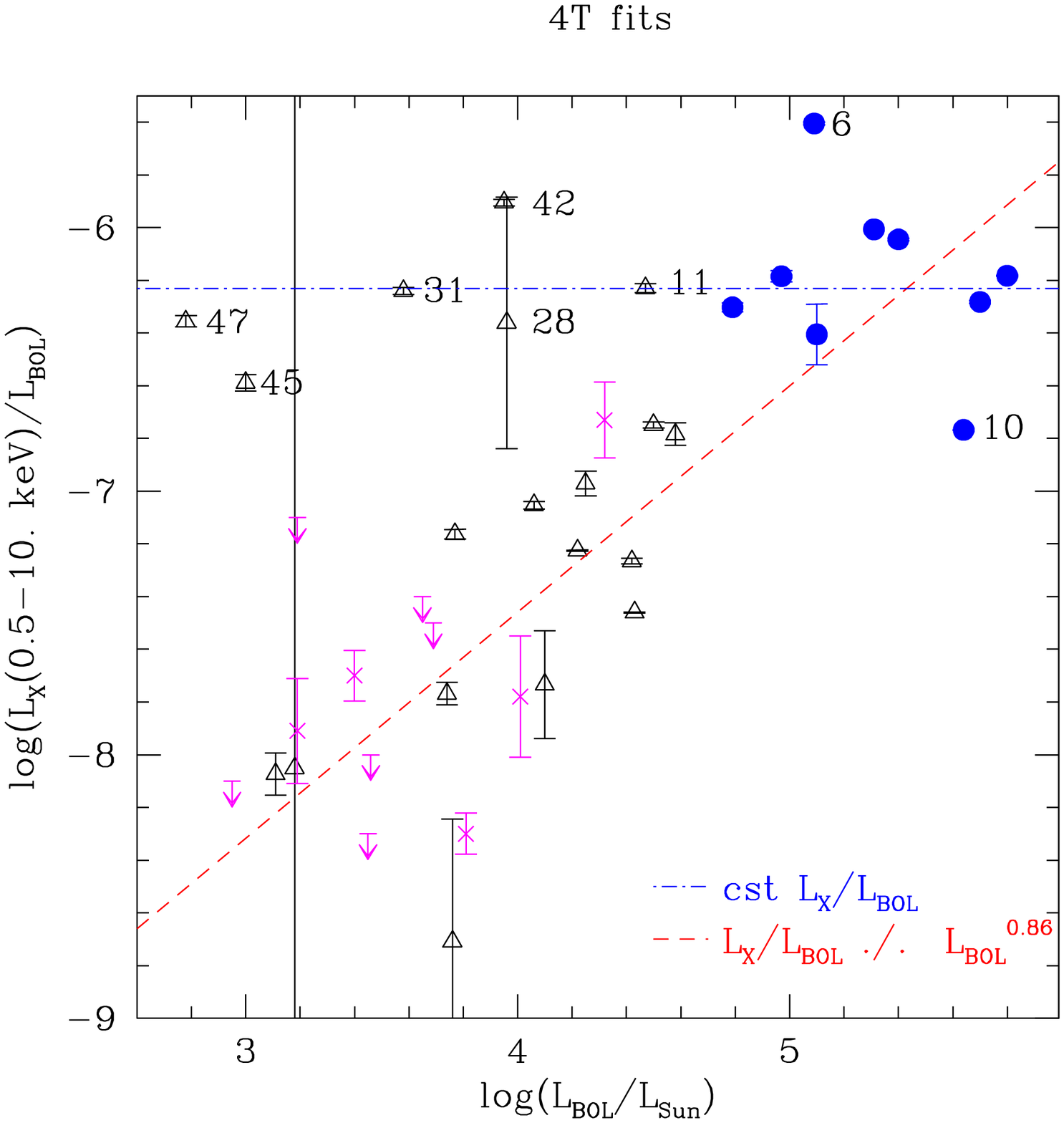}
\caption{Ratio between the X-ray luminosity and the bolometric luminosity as a function of mass-loss rate or bolometric luminosity. Symbols are as in previous figure.}
\label{mdot2}
\end{figure}

To better understand the phenomena at work, we searched for correlations between the X-ray luminosities and the stellar/magnetic parameters, first and foremost the mass-loss rates. Indeed, since every wind shock mechanism extracts kinetic energy from the wind flow and converts it to thermal energy (partly radiated away as X-rays) at shock fronts, we expect some scaling of the X-ray properties with wind mass-loss rate $\dot M$. Fig. \ref{mdot} (top panel) shows that there is of course a strong trend of $L_{\rm X}$ vs. $\dot M$ (from Table \ref{listoftargeta}), albeit with a fair amount of scatter. In particular, a linear relation between $\log(L_{\rm X})$ and $\log(\dot M)$ does not seem to hold for the low/high extreme values of the mass-loss rates. Looking at the \loglxlb\ values (Fig. \ref{mdot2}), we can more clearly distinguish two groups of objects with different behaviours, one with a high value of \loglxlb$\sim-6.2$ notwithstanding the mass-loss rate (group 1) and one with \loglxlb\ varying with $\log(\dot M)$ or $\log(L_{\rm BOL})$ (group 2). 

The first group comprises all of the O stars and six B stars with \loglxlb$>-6.6$ (see IDs in Fig. \ref{mdot}). These objects seem to follow the $L_{\rm X} \propto \dot M^{0.6}$ relation for radiative shocks in high-density winds \citep{owo13} even if three of them (\#31, \#45, and \#47) have very low mass-loss rates. We may also note that two particular cases amongst the O stars, \#6 Plaskett's star (higher luminosity) and \#10 $\zeta$\,Ori (lower luminosity)\footnote{It must be noted that considering or excluding these two stars from the fits do not change the relations reported for group 1.}. The particular properties of these two O stars could be due to the possible non-magnetic origin of their X-rays. For Plaskett's star, not only there is a close companion to the magnetic star but there may be an X-ray bright collision of their stellar winds \citep{lin06}. For $\zeta$\,Ori, the strength of the magnetic field is low, hence its impact on the stellar wind is weak and its X-rays may thus come from embedded wind-shocks, as in non-magnetic stars: the very ``normal'' X-ray properties of this star (no overluminosity, softness of the emission) confirm this view. 

Group 2 consists of the remaining 13 B stars, which follow a significantly steeper trend in mass-loss rate than those in the first group.  

We determined the best-fit power-laws between $\log(L_{\rm X})$ and $\log(\dot M)$ using least squares for these two groups (Fig. \ref{mdot} and Table \ref{relations}). It should be noted that, despite not being included in their derivation, the points corresponding to upper limits and faint detections match well these relations (Figs. \ref{mdot} and \ref{mdot2}). 

The mass-loss rates used for this study were derived from the \citet{vin00} recipe and thus actually correspond to theoretical magnetosphere feeding rates. To use a more ``observational'' value, we have also derived the relations with the bolometric luminosity $\log(L_{\rm BOL})$ (Fig. \ref{mdot2} and Table \ref{relations}). Since the wind is driven by radiation, both types of relations (with $\dot M$ and $L_{\rm BOL}$) yield similar results, but it should be noted that the correlations are slightly tighter for $\log(\dot M)$ than for $\log(L_{\rm BOL})$, which is why we will focus on the former ones in the following. 

We examined whether alternative relations (e.g. with wind density, traced by $\dot M / 4 \pi R_*^2 v_{\infty}$) would yield tighter correlations, but that was not the case. We also examined the relation between the X-ray luminosity and the maximum strength of the H$\alpha$ emission, which should reflect the quantity of material in the magnetosphere, but the latter did not appear to be a good discriminant either\footnote{O stars have similar \loglxlb\ but a large range of H$\alpha$ strengths, whereas B stars have H$\alpha$ equivalent widths close to zero but a large range of \loglxlb\ values}. Finally, we examined relations including additional parameters, such as magnetic field strength or magnetospheric size, e.g. $\log(L_{\rm X})= a \times \log(\dot M)+ b\times \log(B_p) + c$, but they did not result in significantly better fits. This may be due to more complex relations with secondary parameters (see next section) as well as the different ranges covered by the parameters in our sample: terminal velocities $v_{\infty}$ cover a 760--3600\,\kms\ interval (i.e. variations by 0.7\,dex), magnetic field strengths $B_p$ cover a larger 0.2--20\,kG interval corresponding to 2\,dex variations (excluding the extremely low field of $\zeta$\,Ori), while bolometric luminosities $\log(L_{\rm BOL}/L_{\odot})$ cover a range 2.8--5.8 (i.e. 3\,dex variations) and mass-loss rates $\dot M$ are between $10^{-10.4}$ and $10^{-5.5}$\,\msol\,yr$^{-1}$ (or 5\,dex variations). It is thus quite normal for the bolometric luminosity or the mass-loss rate to be the main discriminant for the predicted X-ray luminosities, although the field strength and terminal velocity could explain some of the scatter around the relations shown in Figs. \ref{mdot} and \ref{mdot2}. 

\subsection{Comparison with theoretical predictions}

\begin{figure}
\includegraphics[width=8cm]{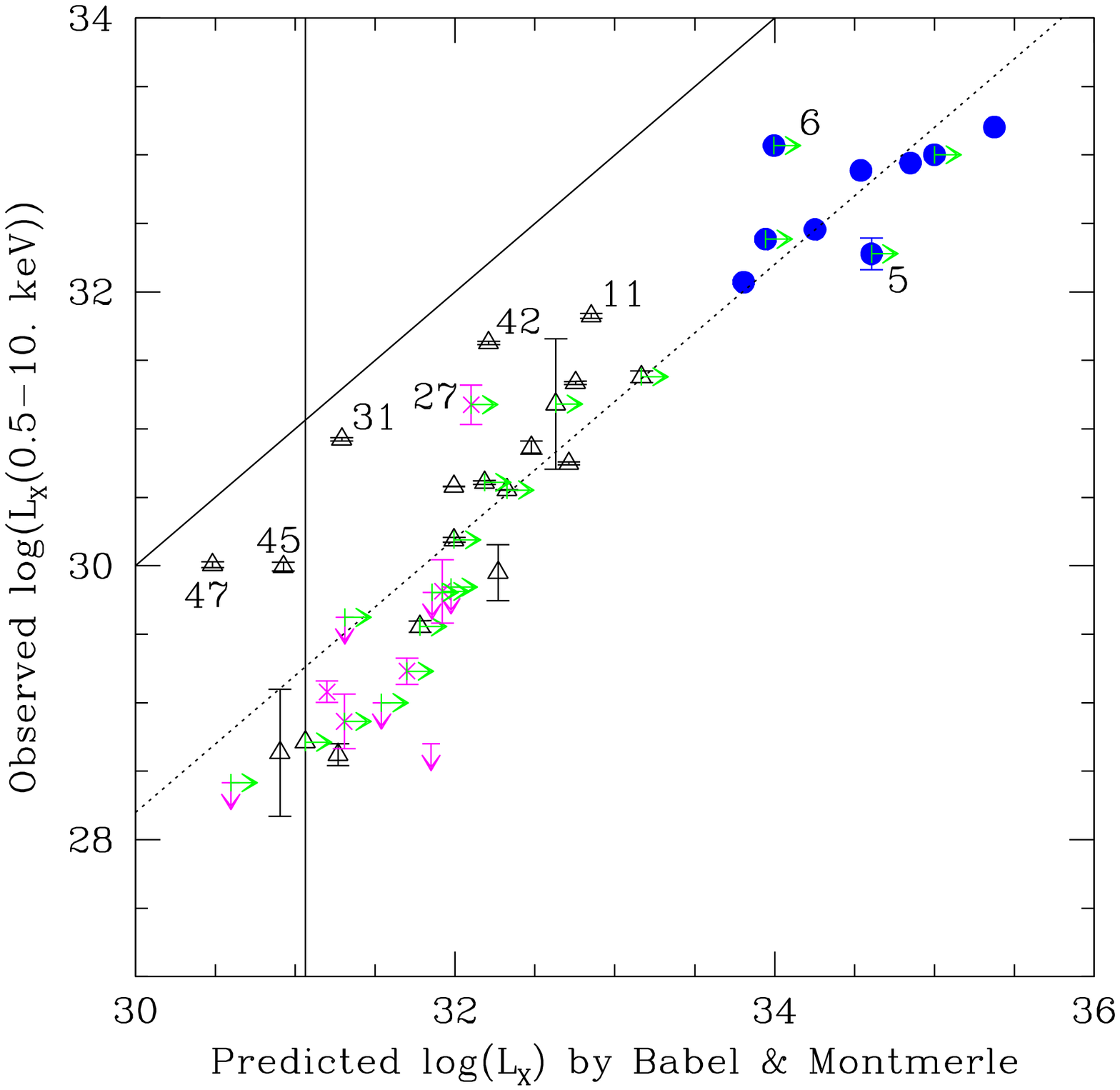}
\caption{Comparison of the X-ray luminosities of magnetic stars (corrected for ISM-absorption, from 4T fits) with the predicted values using the formula of \citet{bab97}. Green rightward-pointing arrows indicate lower limits on the predicted luminosities when only lower limits on the dipolar field strengths are known, the solid line corresponds to a 1-to-1 correlation while the dotted line is located 1.8\,dex below, other symbols are as in Fig. \ref{mdot}.   }
\label{bab}
\end{figure}

The MCW shock model was first discussed in detail by \citet{bab97} on the basis of the confined magnetosphere scenario of \citet{sho90}. While most directly aimed at understanding a single star (IQ\,Aur), the paper of \citet{bab97} also presented a formula to predict the X-ray luminosity of magnetic massive stars with power-law dependences on the three basic parameters magnetic field strength, wind velocity, and mass-loss rate: $L_{\rm X}=2.6\, 10^{30}\, B_*^{0.4}\, \dot M \, v_{\infty}$ where the magnetic field $B_*$ is expressed in units of kG, the mass-loss rate $\dot M$ in units of $10^{-10}$\,\msol\,yr$^{-1}$ and the terminal velocity $v_{\infty}$ in units of 1000\,\kms. Fig. \ref{bab} compares the predictions from this formula (listed in Table \ref{listoftargeta}) with the observed X-ray luminosities in the 0.5--10.0\,keV energy band after correcting for interstellar absorption. The observed and predicted luminosities appear to be globally correlated (confirming the hints of \citealt{osk11}), but two remarks should be made. 

The first and main one is that the Babel \& Montmerle formula fails to reproduce the observed trends. Indeed, a steeper-than-unity slope relates the observed and predicted luminosities for low X-ray luminosities (Fig. \ref{bab}). This is related to our finding of two relations, $L_{\rm X}\propto \dot M^{0.6}$ or $L_{\rm X}\propto \dot M^{1.4}$, rather than one. Second, the predicted values are on average 1.8\,dex higher than the observed ones, for both groups (as illustrated in Fig. \ref{bab}). This may be partly due to the definition of instrumental bandpass (Babel \& Montmerle were analyzing {\it ROSAT} data but never explicitly stated the bandpass to which their formula applies) but also to some idealization of the physics involved (notably a nearly perfect shock heating efficiency).

The pioneering work of Babel \& Montmerle has long been the sole one available, but recently a new study by \citet{udd14} re-investigated the subject, this time considering specific instrumental bandpass and including more precise physics through the use of 2D MHD simulations. They notably found that properly accounting for the post-shock cooling length steepens the relationship at low mass-loss rates through a shock retreat effect\footnote{When channelled winds from opposite footpoints collide near the loop apex, the cooling layer where X-ray emission arises extend from loop top to the shock. In low density plasma, this shock is ``retreated'' back into the accelerating wind, because of the inefficiency of radiative cooling (cooling length comparable to Alfven radius). This effect is thus more pronounced for B stars with low-density winds.}. Additionally, at high $\dot M$, the reduced confinement for a given dipolar strength makes the $L_{\rm X}$ vs. $\dot M$ relation become shallower, and may even turn over at the highest values of the mass-loss rate. These trends are indeed seen in our data.  

\begin{figure*}
\includegraphics[width=9.cm]{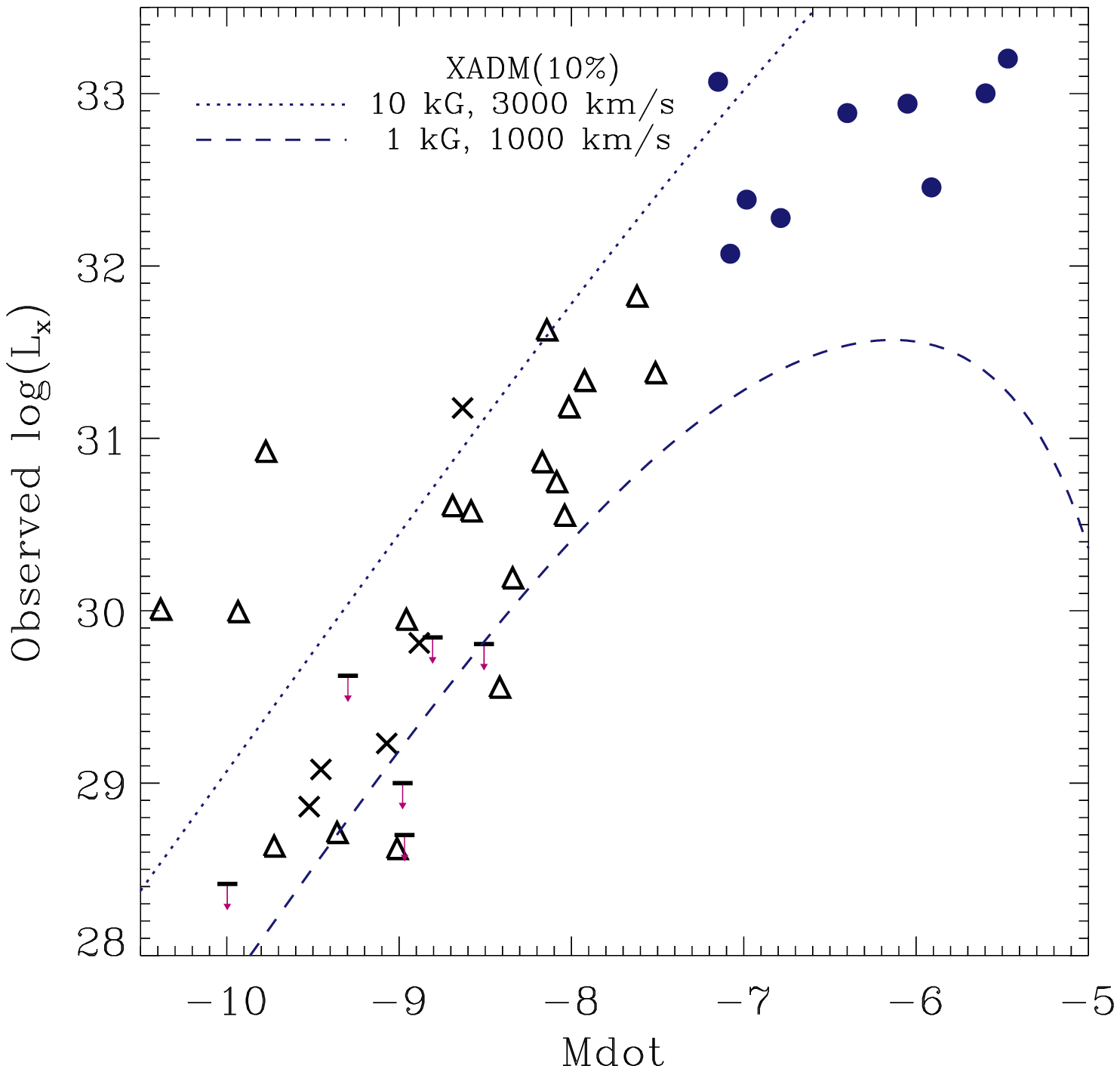}
\includegraphics[width=9.cm]{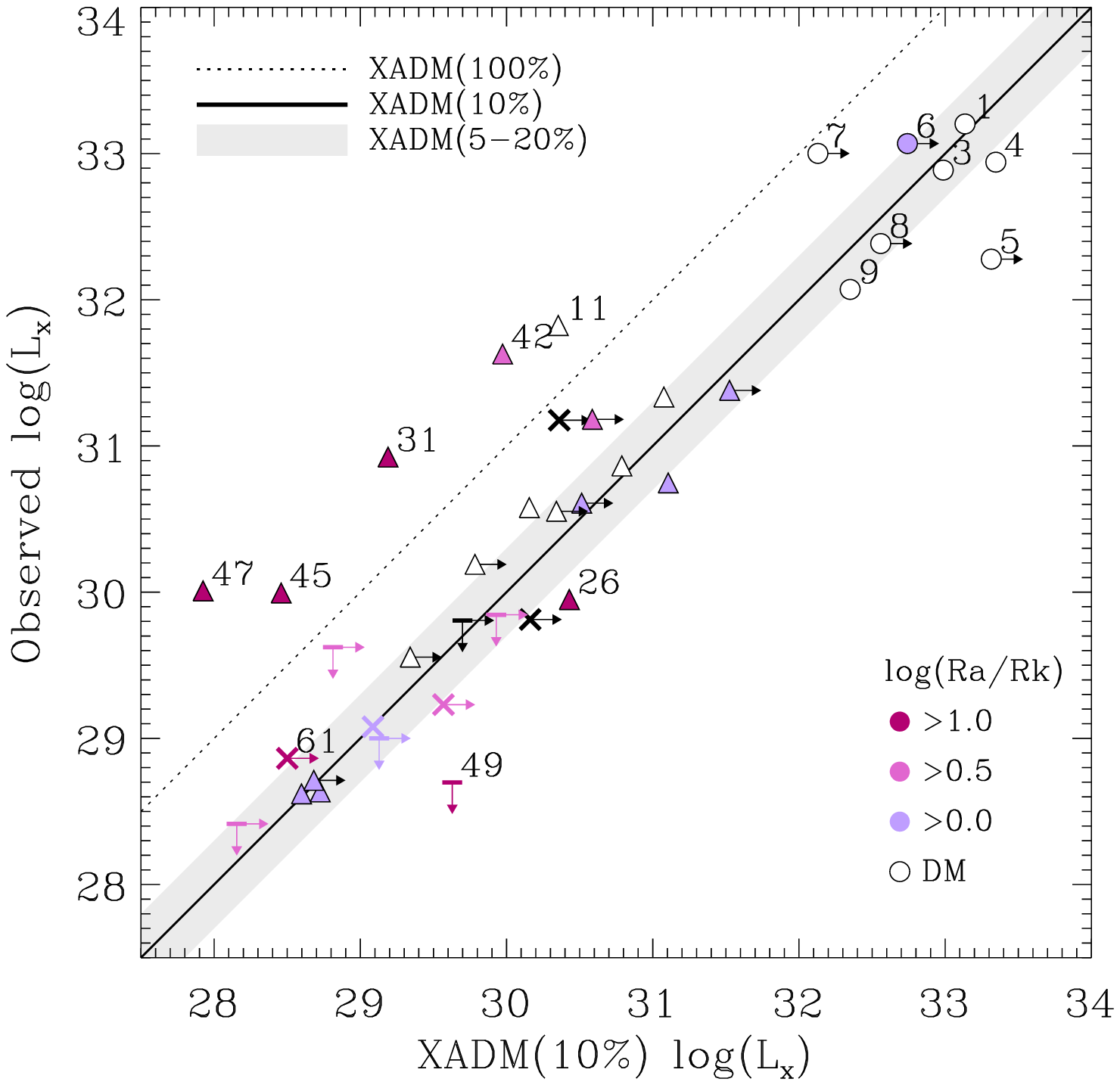}
\caption{{\it Left:} Relation between the X-ray luminosity in the 0.5--10.0\,keV band (corrected for ISM absorption, from 4T fits) and logarithm of mass-loss rate. The dotted and dashed lines represents predicted X-ray luminosities in the same energy band using the XADM model \citep{udd14}, scaled by 10\%, for a single value of the stellar radius of 10$^{12}$\,cm and two set of (indicated) magnetic field and wind parameters bracketing the parameters of our sample. Symbols are as in Fig. \ref{mdot}. {\it Right:} Comparison between the observed X-ray luminosity of magnetic stars (as in left panel) and the predicted values using the XADM model of \citet{udd14}. The dotted line illustrates the ideal model with 100\% efficiency whereas the solid line indicates a scaling by 10\%; the grey shaded area corresponds to scalings by 5--20\% (a range in efficiency consistent with MHD models). Rightward-pointing arrows indicate stars for which the dipolar field strength is a lower limit. The symbols are colour-coded according to the predicted size of their centrifugally supported regions caused by rapid rotation (with darker shades for larger sizes); stars without predicted centrifugal support (see \citealt{pet13}) have empty symbols. Stars of particular interest are labeled according to their identification number in Table \ref{listoftargeta}. Note that $\zeta$\,Ori (\#10) is not present on this plot as the XADM model is only valid for stars with significant magnetic confinement (i.e., $\eta_*>1$). }
\label{xadm}
\end{figure*}

As MHD simulations are computationally costly, and nearly impossible for $\eta_* \sim 10^4-10^6$ confinement values appropriate for many observed magnetic B stars, \citet{udd14} developed a semi-analytic ``XADM'' model, which predicts the X-ray luminosities as a function of magnetic field strength, stellar radius, and wind velocity. It reproduces well the overall trends in the MHD-computed $L_{\rm X}$ vs. $\dot M$ but the idealized XADM analysis, which assumes 100\% efficiency, display larger values than the MHD results. This reflects a reduced X-ray efficiency ($\lesssim$20\%) from infall and other dynamical effects in the full 2D MHD simulations, implying that a scaling of the XADM model is necessary. The left panel of Fig. \ref{xadm} shows that predictions by this model, scaled by 10\% and for parameters bracketing the most extreme combinations of magnetic field and wind velocity in our sample, bracket well the observed data. To explore the effects of additional model parameters, we computed the X-ray luminosities using the individual parameters of our targets (see last column of Table \ref{listoftargeta} and the right panel of Fig. \ref{xadm}): a good agreement is found, with less scatter than in Figs. \ref{mdot} and \ref{bab}.

The overall observed emission levels suggest efficiencies in the range of 5--20\% (grey area in Fig. \ref{xadm}), with a best-fit around 10\%. Note that refining this value will notably require more precise determinations of the mass-loss rates since both a lower $\dot M$  and a lower efficiency similarly lead to a lower X-ray emission. Indeed, detailed studies \citep[e.g.][]{osk11} often show significantly lower mass-loss rates than theory predicts and mass-loss rates of OB stars, generally inferred from density-squared diagnostics, could be lowered by a factor of $\sim 3$ to take into account the effects of wind clumping \citep{osk08}: the theoretical mass-loss rates used here may thus have to be revised downward. 

\begin{figure*}
\includegraphics[width=8cm]{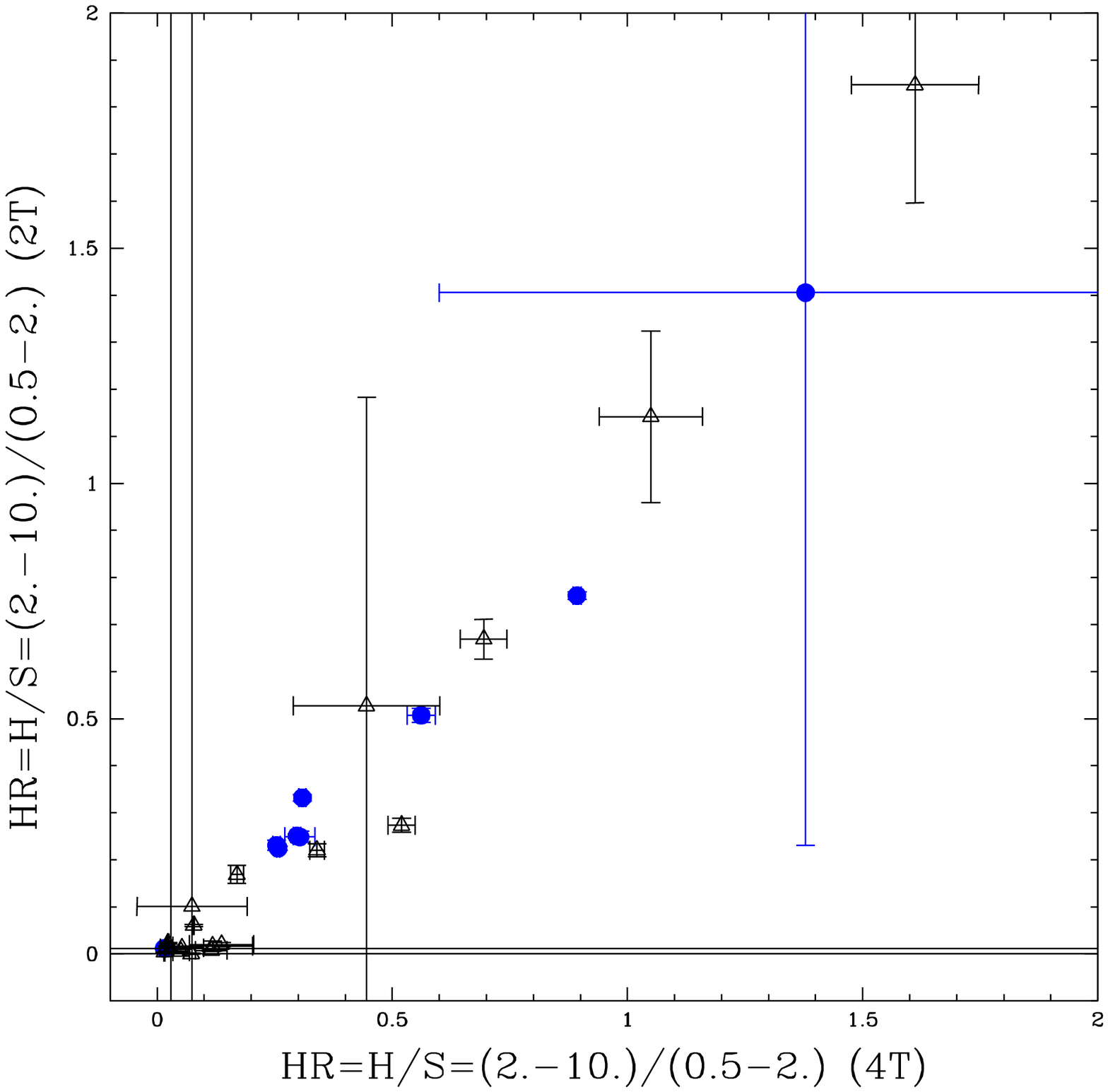}
\includegraphics[width=8cm]{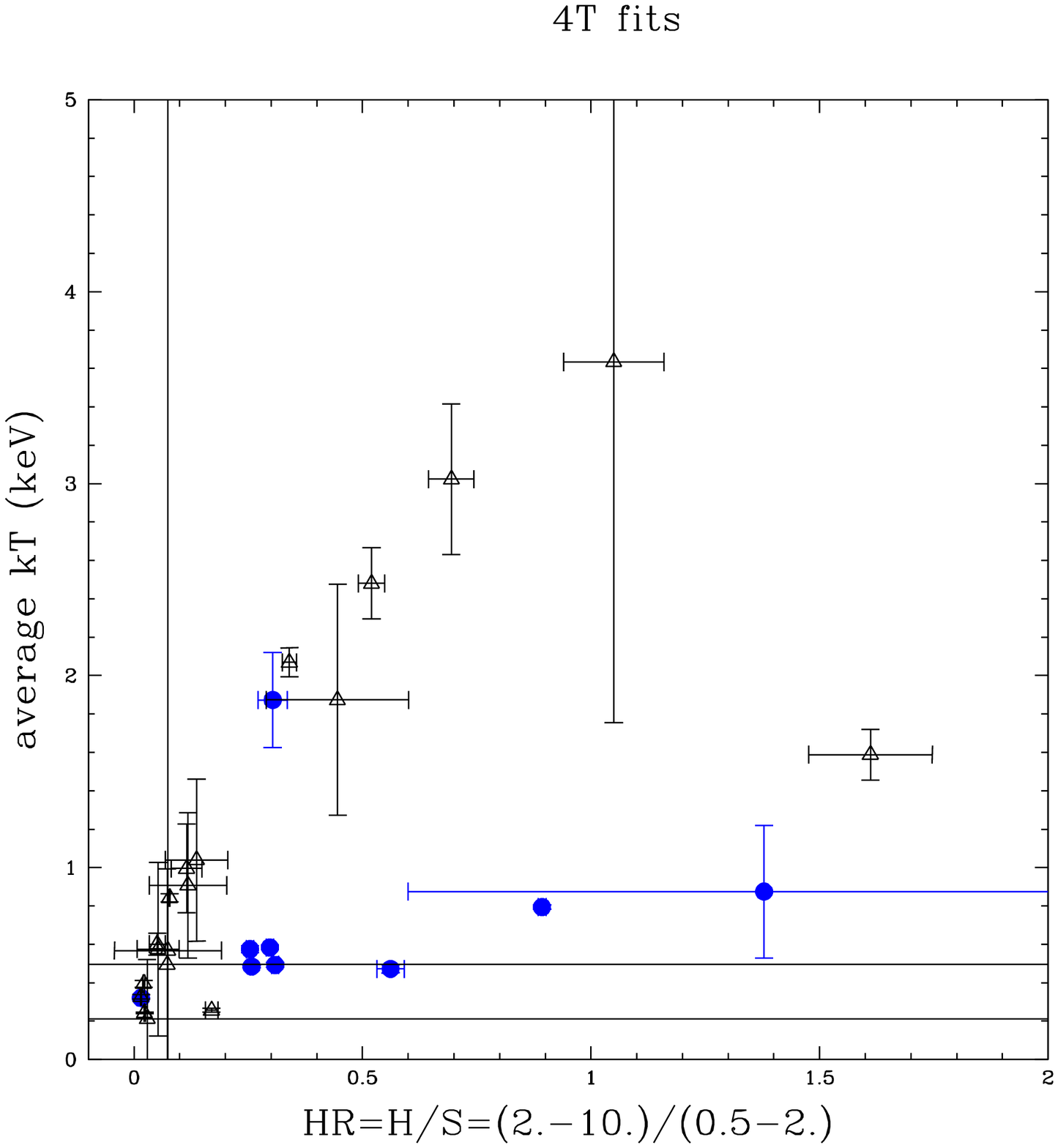}
\caption{{\it Left:} Comparisons of the hardness ratios obtained with the two fitting choices: they are independent of the method, showing their robustness. {\it Right:} Relation between hardness ratios and temperatures (4T fits). There is no unique linear relationship between the two parameters, but an interesting dichotomy showing that two values of the same average temperature yield the same hardness ratio.  Filled blue dots correspond to O stars, black empty triangles to B stars.}
\label{compamethod}
\end{figure*}

A few discrepancies nevertheless exist between data and the predictions from \citet{udd14}, as can be noted in Fig. \ref{xadm} by the presence of stars with over- or under-luminosities. Amongst O stars, the most discrepant point is NGC1624-2 (\#5) which displays an X-ray luminosity lower by about 1\,dex compared to the model prediction (see also Fig. \ref{bab}). In this context, it must be recalled that the model does not take into account the absorption of the X-ray emission by the cool wind material. However, the spectra of the O stars require such additional absorption to be reproduced by $apec$ thermal models (see Sect. 4.2), which is not surprising in view of their dense stellar winds. Therefore, the amount of X-rays actually generated, which should be compared with the model, may be higher than the emergent X-ray emission level derived here. This effect, which may explain why the efficiency appears slightly lower for O stars than for B stars (Fig. \ref{xadm}), particularly applies to NGC1624-2 (\#5) as this star hosts the largest magnetic field amongst O stars, hence the largest magnetosphere, and requires the highest absorption ($1.2\times10^{22}$\,cm$^{-2}$, see Tables \ref{4tfits} and \ref{2tfits}). Its generated X-ray emission level could easily be 1\,dex higher (see Petit et al., submitted). The other apparently deviant point amongst O stars is HD\,108 (\#7), which appears brighter than expected (Fig. \ref{xadm}). However, the detailed magnetic properties of this object are not known, and the presence of an overluminosity can thus only be truly ascertained after further monitoring.

There are also five B-type stars with observed luminosities larger by at least one dex compared to the predictions (Fig. \ref{xadm}): those are \#11 $\tau$\,Sco, \#31 $\sigma$\,Ori\,E, \#42 HD\,200775, \#45 HD\,182180, and \#47 HD\,142184. This is unlikely to be due to the contamination by low-mass companions undergoing X-ray bright flares: the observed X-ray emission is mostly soft, and no flares are detected in their lightcurve. Another possible origin for this discrepancy is the uncertainty in mass-loss rate, as the X-ray luminosity is a strong function of $\dot M$. Beyond remarks already made above, this particularly applies to stars near the bi-stability jump (22.5$<T_{eff}<$30\,kK), but only one object out of 5 belong to this category. A last possibility for explaining the high luminosity of these stars is the role of rapid stellar rotation, which can contribute significant centrifugal acceleration to the pre-shock wind plasma \citep{udd08,tow05} but is not included in the models of \citet{udd14}. However, while four of these five B stars are among the rapid rotators with the most extreme magnetospheres (darker symbols in the right panel of Fig. \ref{xadm}), it is not a general rule, as the luminosity of some other rapidly-rotating objects are well reproduced by the model (e.g. \#26, \#49, or \#61) and one overluminous star (i.e. \#11) is not rapidly rotating. In the same vein, a mismatch also exists for $\tau$\,Sco (\#11) whose high-energy behaviour differs from those of its ``clones'' \citep[\#13 HD\,63425 and \#14 HD\,66665, ][and this work]{ign13}. Further investigation of the effect of rotation, combined with multi-wavelength detailed study of the individual stars to mitigate the mass-loss rate uncertainties would thus be useful. 

\section{Other X-ray properties}
\subsection{Hardness or temperature}

\begin{figure}
\includegraphics[width=8.5cm,bb=30 150 420 700, clip]{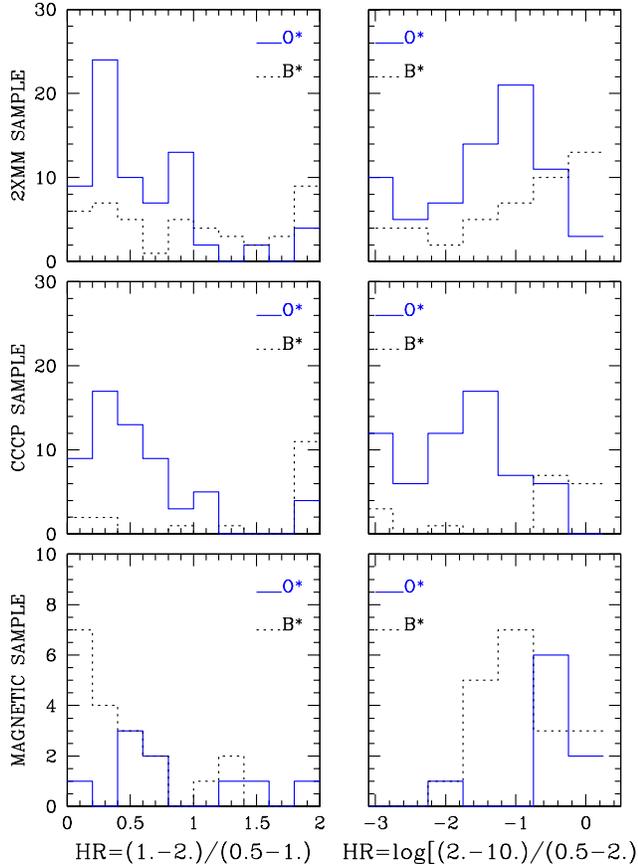}
\caption{Hardness ratios of massive stars in 2XMM \citep[top panels]{naz09} and CCCP \citep[middle panels]{naz11} compared to values for stars in this paper (bottom panels). Note that one limit of the energy bands is different (2\,keV in our data, 2.5\,keV in CCCP and 2XMM).}
\label{histohr}
\end{figure}

\begin{figure}
\includegraphics[width=8cm]{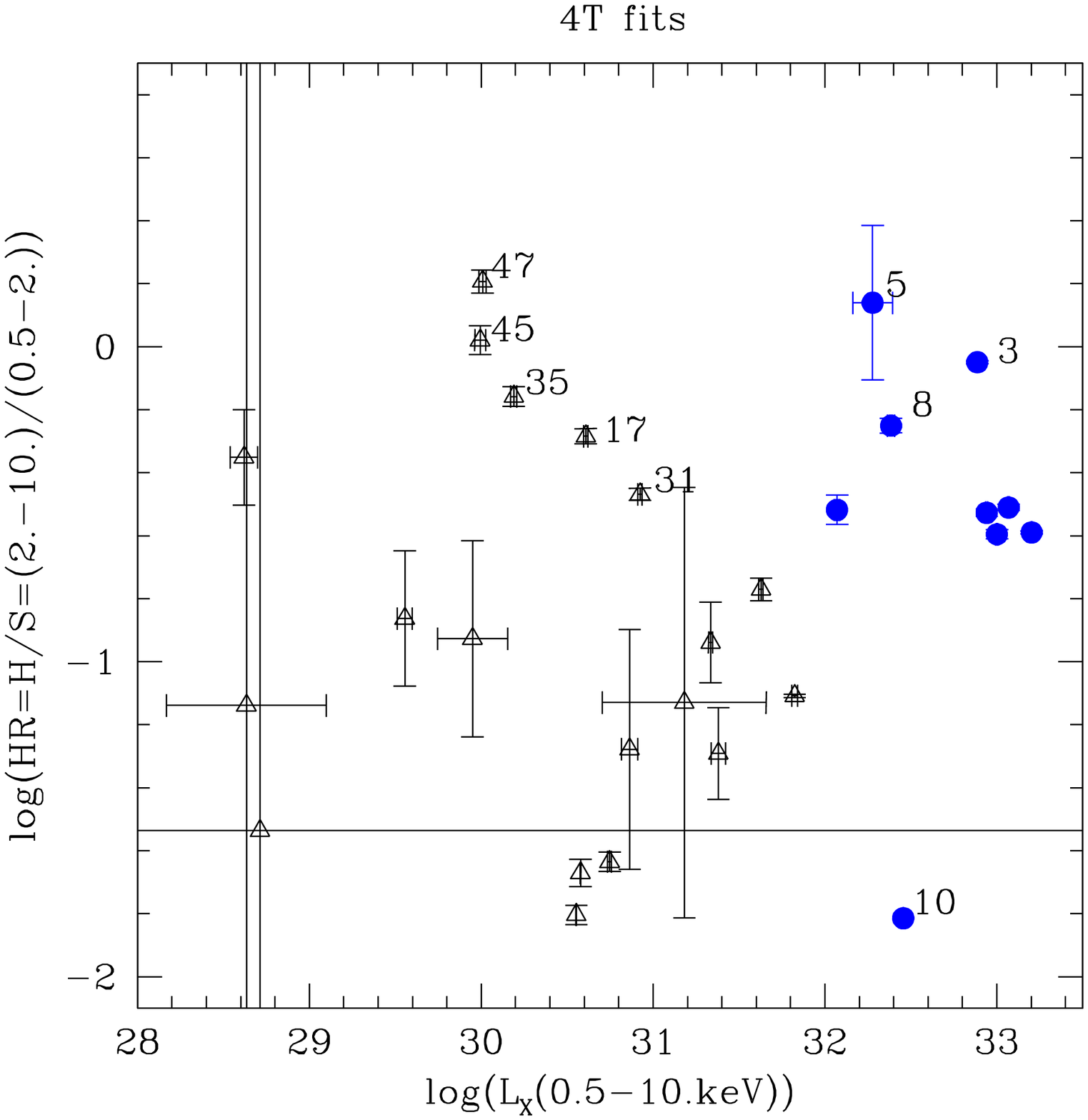}
\caption{Relation between hardness ratios and X-ray luminosities. Symbols are as in Fig. \ref{compamethod}.}
\label{hrobs}
\end{figure}

Our fittings also characterize the spectral shape of the X-ray emission, through two parameters. The first one is a ``hardness ratio'' calculated as the ratio between the hard (2.--10.0\,keV) and soft (0.5--2.0\,keV) ISM absorption-corrected fluxes. This hardness ratio appears very reliable as it is notably independent of details of the fitting itself (see left panel of Fig. \ref{compamethod}). However, when the source is both faint and relatively soft, the hard X-ray flux (hence the hardness ratio) may be underestimated. It is a problem which is difficult to solve without better observations, but it should be noted that it does not prevent us from detecting sources mostly composed of very hot plasma. The second parameter is the average temperature, defined as $\overline{kT} = (\sum kT_i \times norm_i)/(\sum norm_i)$, where $norm_i$ are the normalization factors of the spectral fits, i.e. $10^{-14} \int n_e n_{\rm H}dV/4\pi d^2$, listed in Tables \ref{4tfits} and \ref{2tfits}. This parameter is sometimes used as diagnostic for the properties of the X-ray emission of massive stars \citep{gag11,ign13}. However, we found it to be not as reliable as the hardness ratio (Fig. \ref{compamethod}). Indeed, even for bright sources, there is a trade-off between absorption and temperature, which means that a unique spectral shape can be fitted by several solutions with very different average temperatures. This effect was reported several times in the past \citep[e.g.][]{naz07}, even when high-resolution spectra were used \citep{naz12}. In this work, we were regularly confronted with this problem, e.g. fits with similar hardness ratios but average temperatures of 0.2 and 1.0\,keV fit equally well the spectra of $\sigma$\,Ori\,E. The main problem is that it is difficult to securely constrain independently the level of additional absorption: average temperatures should thus not be taken at face value. Fortunately, the same conclusions are reached for both parameters, so that we present only hardness ratios in the following.

\begin{figure}
\includegraphics[width=8.5cm]{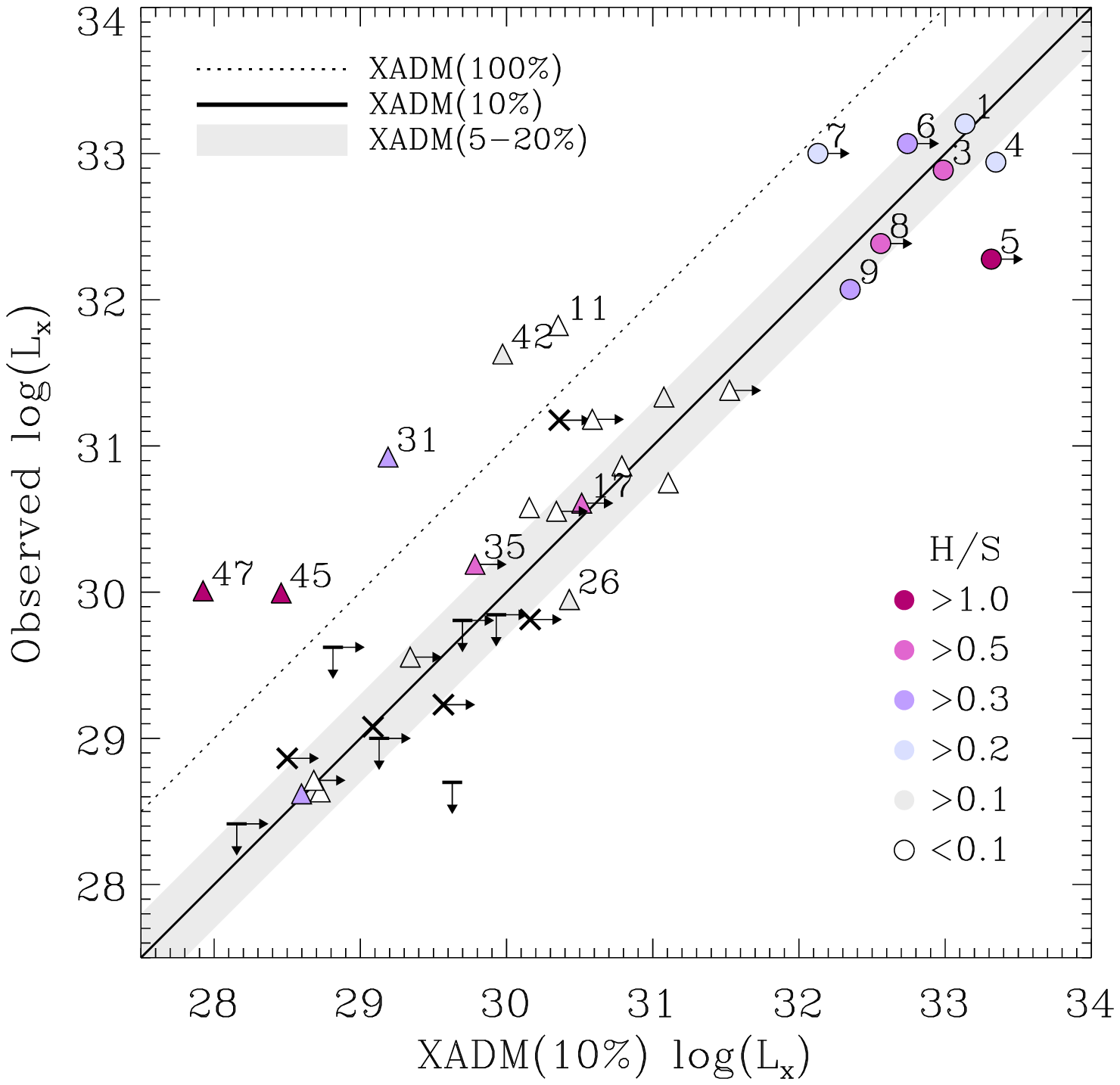}
\caption{Same as right panel of Fig. \ref{xadm} with symbols colour-coded according to their measured hardness ratios. No strong trend emerges, though it is interesting to note that three overluminous B stars, which are rapid rotators, display hard X-rays. However, as before, other rapidly-rotating B stars do not stand out: while the connection with rapid rotation is suggestive, it is certainly far from conclusive. }
\label{hrtheo}
\end{figure}

Our targets cover a wide (2\,dex) range of hardness ratios (see e.g. Table \ref{observable}). Compared to large samples of generally non-magnetic stars (Fig. \ref{histohr}), the magnetic O stars appear to have larger hardness ratios. Indeed, $\zeta$\,Ori has the lowest hardness ratio of our sample, consistent with the probably non-magnetic origin of its X-rays. The situation is different for B stars: the magnetic B stars of our sample appear softer than most B stars in CCCP or 2XMM. However, as noted above, the CCCP and 2XMM sample mostly X-ray bright B stars, which may introduce an observational bias. 

To assess the possibility of a link between the spectral shape and the emission level, Fig. \ref{hrobs} shows hardness ratios as a function of X-ray luminosities. The situation appears quite varied. O stars display various hardness ratios despite similar X-ray luminosities (and even more similar \lxlb\ ratios). Most of the B stars display low ($<0.2$) hardness ratios despite an extended range of luminosities though there are also a few objects displaying elevated hardness ratios compared to other stars of similar luminosities. Amongst them, we find three B stars with brighter than predicted X-ray emission (\#31, \#45, and \#47) as well as two B stars for which observation and prediction agree (\#17 and \#35).  

We searched for correlations between hardness/temperature and the stellar or magnetic parameters but none are found. From the theoretical point of view, the MCW shock mechanism should, in principle, result in higher plasma temperatures, because of the higher velocity jump in the head-on MCW shocks compared to those in stochastic embedded wind shocks. It is true that the emission of magnetic O stars appears, on average, slightly harder than that of ``normal'' O stars, but the scatter in each group is large. In addition, the situation is opposite for magnetic B stars, whose X-ray emission is predominantly soft, even slightly softer than ``normal'' B stars. Detailed models by \citet{udd14} further predict a harder X-ray emission for a higher confinement and/or a higher mass-loss rate but this theoretical prediction is not verified. Also, there is no obvious link between the detected overluminosities and hardness ratios (Fig. \ref{hrtheo}). These facts, together with the large scatter amongst O stars, suggest that the main driver for plasma temperature properties must be different than the main driver for the X-ray luminosity, which may guide future theoretical modelling.

\subsection{Absorption}

Another interesting observable to investigate is the additional, local absorption (i.e. above interstellar absorption). For ``normal'' O stars, the high-energy emission arises throughout the wind, whose cool parts may absorb X-rays. Some absorption effects are indeed detected, in the shapes of the line profiles \citep{coh10,her13} as well as in the additional absorption needed to model spectra \citep[with values up to $8\times10^{21}$\,cm$^{-2}$ and an average of $4\times10^{21}$\,cm$^{-2}$, see e.g.][]{naz09,naz11}. No such additional absorption is needed to model the spectra of ``normal'' B stars \citep{naz09}. However, as magnetic stars possess dense magnetospheres, significant local absorption might still be possible. 

Except for the highest value (found in NGC1624-2, the most magnetic O star), the absorption values found for magnetic O stars are not significantly different from what is observed in ``normal'' O stars. It may be noted that the O stars with the three highest absorptions also display the three highest hardness ratios, pointing to absorption as responsible for elevated ratios, rather than temperature (Fig. \ref{hrobs} and Tables \ref{4tfits}, \ref{2tfits}). However, no correlation between absorption and the stellar/magnetic parameters could be found. 

In a similar way, magnetic B stars do not generally require additional absorption, like their non-magnetic siblings. Whenever best-fits require additional absorptions (Tables \ref{4tfits} and \ref{2tfits}), a fit with (forced) zero additional absorption has a similar quality. There are however two exceptions (\#29 and \#42). It is difficult to find a common point between these two objects, apart from the fact that they have both been reported to be Herbig stars - but they are not the only ones in our sample. Rapid rotation, associated with large and dense magnetospheres, is not an explanation either, as the latter object is rapidly-rotating, but not the former, and other rapidly-rotating magnetic stars do not show such increased absorption.

\subsection{Variability of the X-ray emission}
 
To assess the presence of changes in the observed X-ray emission, several exposures are needed but in our dataset, only eight targets have more than one spectrum available. We have therefore examined this subsample, by searching for variability between the individual exposures (see Table \ref{variabsum} for a summary). It should be mentioned that this subsample represent only a small fraction of the Petit et al. catalog: the derived conclusions thus do not have a general character. 

\begin{figure*}
\includegraphics[width=15cm, bb=25 150 520 490, clip]{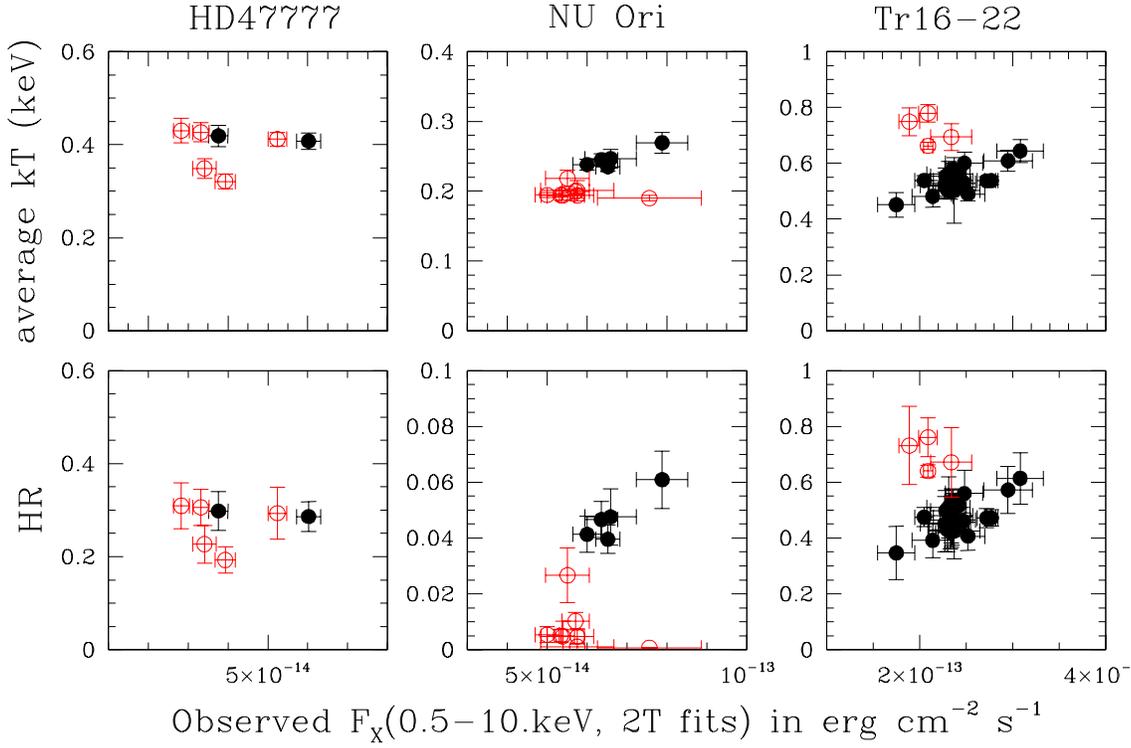}
\caption{Evolution of the average temperatures and hardness ratios ($HR=H/S=F^{ISMcor}(2.-10.0\,keV)/F^{ISMcor}(0.5-2.\,keV)$) as a function of observed fluxes for the three objects with many ($>6$) observations: HD\,47777 (left), NU\,Ori (middle) and Tr16-22 (right). Red open symbols correspond to \ch\ data, while black filled symbols correspond to results from \xmm\ observations. The values shown here correspond to the results from 2T fits (results are similar for the 4T fits, and for ISM-absorption corrected fluxes). HD\,47777 displays flux variations but no large spectral changes, whereas NU\,Ori and Tr16-22 display simultaneous flux and hardness variations. }
\label{variab}
\end{figure*}

\subsubsection{Stable X-ray emission: HD\,148937, $\sigma$\,Ori\,E, and $\zeta$\,Ori}

The observations (\xmm\ only) of $\zeta$\,Ori yield remarkably stable results. For HD\,148937, there is also no significant difference between the two \xmm\ observations, but the average temperature appears 10\% higher in the \ch\ exposure, and the associated flux are about 15\% lower. This is reminiscent of the results of \citet{sch13} for isolated galaxy clusters. The detected changes therefore seem to be related to cross-calibration differences and cannot be attributed to the star without further information. A similar situation is found for $\sigma$\,Ori\,E.

The constancy of the X-ray emission of these three objects can be attributed to their magnetic and geometric properties. HD\,148937 has a low obliquity angle between the magnetic pole and the rotational axis, so that the variations of the line profiles in the optical domain are of very small amplitude \citep{naz10,wad148937}. This geometry also reduces the possibilities of change in occultation or absorption (see Sect. 4.3.3), explaining the stable X-ray emission. For $\sigma$\,Ori\,E, the rotation period is short (1.19d, or 103ks) so that changes related to confined winds would be easily detected during a long exposure like the \ch\ observation (91ks) but that is not the case \citep[ see in particular their Fig. 5]{ski08}. This absence of phase-locked X-ray variations in $\sigma$\,Ori\,E explains the stability of the X-ray emission when comparing \xmm\ and \ch\ datasets. It could be related to the large size of the magnetosphere (about 31R$_*$ for the Alfven radius), which minimizes the impact of occultations in the high-energy domain. Finally, the wind confinement of $\zeta$\,Ori is very small, so that the wind is nearly undisturbed by the presence of the (small) magnetic field. Its X-ray emission, most probably linked to embedded wind shocks as for non-magnetic O stars, is not expected to vary \citep[cf. $\zeta$\, Pup, ][]{naz13zeta}.

\subsubsection{Flux variations without obvious spectral changes: $\beta$\,Cep and HD\,47777}

The observations of $\beta$\,Cep display small flux variations, though the spectral shape (hence the average temperature and hardness ratio) appears stable. This confirms the analysis of \citet{fav09} of the same data. The available datasets were taken over 10d, an interval comparable to the rotation period (12d), but, without additional data, it is difficult to judge whether these small flux variations are truly correlated with the stellar rotation period: their origin remains uncertain.

HD\,47777 also displays a quite stable spectral shape, but its flux varies by at least a factor of two (Fig. \ref{variab}). A rotation period of 2.64d (about 228ks) was recently derived for this object \citep{fos14}. The longest available exposures cover about half of that period (Table \ref{listoftargetb}), and we therefore examined the X-ray lightcurves to search for variations compatible with it. None was detected, but $\sim$10ks flares at least tripling the X-ray emission level are observed in \ch\ exposure \#2540 and in \xmm\ exposure \#0011420101. Outside flaring episodes, the flux appears less variable. It must be recalled that HD\,47777 is a Herbig star, hence the presence of flares is not totally surprising (e.g. through the possible contamination by X-rays from a low-mass companion). Their link with the MCWs is certainly not established.

\subsubsection{Flux variations with spectral changes: HD\,191612, NU\,Ori, Tr16-22}

The changes in HD\,191612 were reported by \citet{naz07,naz10}, and we confirm here the results of varying flux and spectral shape. These variations are clearly phase-locked, with the maximum (resp. minimum) flux occurring when the dense equatorial regions are seen face-on (resp. edge-on) \citep{naz10}. 

Similar variations are seen for NU Ori (Fig. \ref{variab}). In addition, the \ch\ data systematically yield fittings with lower hardness ratios and lower average temperatures, which is contrary to known calibration effects \citep{sch13}. This suggests that the target truly was softer when observed with \ch.

Tr16-22 was observed more than 20 times in the X-ray range and shows clear flux and spectral shape variations (Fig. \ref{variab}, and \citealt{naznew}). Changes are best seen on \xmm\ data, because of their higher quality, their higher number, and the fact that they cover a large range of flux and hardness values. The large number of exposures helped us identifying, for the first time, the possible period of Tr16-22 \citep{naznew}.

Variations in both flux and spectral shape are thus detected for three stars, as well as for $\theta^1$\,Ori\,C \citep{gag05, ste05}. Moreover, for HD\,191612, $\theta^1$\,Ori\,C, and possibly Tr16-22\footnote{The period of NU\,Ori is unknown, so that we cannot check the relation with stellar rotation.}, the flux variations are clearly phased with the stellar rotation period, as demonstrated by the simultaneous minima in X-ray and optical emissions. This can be qualitatively explained in the case of magnetic oblique rotators: as the view on the magnetosphere changes as the star rotates, occultation of the X-ray emitting regions by the star or obscuration by its confined wind can cause periodic variations, recurring with the stellar rotation period. However, the latter explanation can be discarded: a large increase in absorption would be required to explain the observed flux decreases, but no significant one (i.e. larger than 2--3$\sigma$) was detected during our spectral fitting nor for $\theta^1$\,Ori\,C, \citep[see in particular their Fig. 14]{gag05}. Moreover, if absorption was the cause of the flux variations, the X-ray emission would actually always be softer when the flux is minimum, which is not the case (see below).

We are thus left with the hypothesis of occultation as the main cause of the X-ray variations. In this case, the observation of simultaneous flux and hardness changes indicates that the X-ray production region is somewhat stratified in temperature. However, HD\,191612, Tr16-22, and NU\,Ori, appear harder when brighter, while $\theta^1$\,Ori\,C appears to be softer while brighter in the COUP data \citep[see Fig. 7 in ][]{ste05}, i.e. the opposite behaviour despite the fact that $\theta^1$\,Ori\,C is not conceptually different from the three others. Since there are two hardness/flux trends, the inferred magnetospheric geometry must differ from star to star: for $\theta^1$\,Ori\,C, the warm plasma, responsible for the soft X-rays, is the one suffering most from occultation and should thus be confined in a smaller region; for the other stars, the hard X-rays should be produced closer to the photosphere and magnetic equator, explaining their disappearance during occultations, while the soft X-rays should be produced in a larger region, less prone to such effects. 

Up to now, the problem of flux or hardness changes was not addressed in published MCW models yet. The behaviours reported above therefore represents strong observational constraints, that can help guiding future modelling. 

\section{Conclusion}

The goal of this analysis is to study the X-ray emission of magnetic OB stars. With this aim, we have analyzed a large series of X-ray observations (more than 100 exposures), which cover $\sim$60\% of the known magnetic massive stars listed recently by \citet{pet13}. Spectra were extracted and fitted by thermal models whenever the sources were bright enough (28 cases). In addition, we converted the count rates into X-ray luminosities for 5 faint sources, while upper limits on the X-ray luminosities were derived for 6 undetected objects. Two O stars ($\zeta$\,Ori and Plaskett's star) show distinct X-ray properties, indicating that the main origin of their X-ray emission is most probably non-magnetic.

We analyzed the whole sample in quest of relations between X-ray properties and stellar/magnetic parameters. The X-ray luminosities follow $L_{\rm X} \propto \dot M^{0.6}$ (hence \loglxlb\ is constant at $-6.2$) for O stars and a few B stars, and $L_{\rm X} \propto \dot M^{1.4}$ for most B stars. Considering alternative relations or a two-parameter dependence of the luminosity, e.g. on mass-loss rate and magnetic field rather than on mass-loss rate only, does not improve significantly the results. It must be noted that the observed X-ray luminosities, and their trends with mass-loss rates, cannot be explained within the framework of embedded wind shocks. On the other hand, luminosity predictions using the \citet{bab97} model of MCWs are too high (by 1.8\,dex) and fail to reproduce the observed trends at high and low mass-loss rates. New MHD modelling of MCWs including shock retreat \citep{udd14} are however able to match observations (level of X-ray emission, trend with mass-loss rates) fairly well. There are nevertheless five B stars much more luminous than expected: most of them are rapid rotators, but not all; in a similar way, not all rapid rotators are overluminous. The origin of their intense X-ray emission thus remains uncertain, requiring more observational and theoretical work.

Regarding other observables, the situation appears less clear. Additional absorption is needed for fitting X-ray spectra from O stars and a few B stars, but no obvious correlation with stellar/magnetic properties is detected. Spectral shape varies amongst the targets, but again, no obvious correlation with stellar/magnetic properties is detected, contrary to expectations of harder emission with higher confinement or mass-loss rate. Further theoretical work is thus required to identify the missing ingredient explaining this (lack of) trends.

Finally, when several observations were available, we have examined the variability of the objects and observe three different behaviours. First, X-ray characteristics were found to be constant in some cases, as expected from the properties of the target (pole-on geometry for HD\,148937, large magnetosphere for $\sigma$\,Ori\,E, non-magnetic origin of the X-rays for $\zeta$\,Ori). Second, flux changes without changes in spectral shape are observed in two cases (HD\,47777 and $\beta$\,Cep) but they cannot be linked to the confined wind phenomenon using current data. Lastly, periodic changes in flux and spectral shape are also observed (e.g. HD\,191612 and $\theta^1$\,Ori C). They are most probably linked to occultation effects in magnetic oblique rotator systems and suggest various temperature stratification for the MCW regions.

\acknowledgments
\section*{Acknowledgments}
YN acknowledges careful reading by G. Rauw, as well as support from  the Fonds National de la Recherche Scientifique (Belgium), the Communaut\'e Fran\c caise de Belgique, the PRODEX XMM and Integral contracts, and the `Action de Recherche Concert\'ee' (CFWB-Acad\'emie Wallonie Europe). VP acknowledges support from NASA through Chandra Award numbers G02-13014X and TM-15001C issued by the Chandra X-ray Observatory Center which is operated by the Smithsonian Astrophysical Observatory for and behalf of NASA under contract NAS8-03060. MR acknowledges support from NASA ATP Grant NNX11AC40G and from University of Delaware. AuD thanks NASA for support through Chandra Award number TM4-15001A. GAW acknowledge Discovery Grant support from the Natural Science and Engineering Research Council of Canada (NSERC). ADS and CDS were used for preparing this document. 

{\it Facilities:} \facility{XMM}, \facility{CXO}.

\clearpage

\begin{landscape}
\begin{table}
\footnotesize
  \caption{ONLINE MATERIAL -- List of targets, with their stellar and magnetic properties.}
  \label{listoftargeta}
  \begin{tabular}{llcccccccccccc}
  \hline
ID & Name  & sp. type & $N_{\rm H}$(ISM) & $d$ & $\log(L_{\rm BOL}/L_{\odot})$ & $R_A$ & $R_K$ & $\log(\dot M)^a$ & $v_{\infty}$ & $R_*$ & $B_p$  & $\log(L_{\rm X})$(B\&M97) & $\log(L_{\rm X})$(udd14)$^b$\\
 &  &  & (10$^{22}$\,cm$^{-2}$) & (pc) & & (R$_{\odot}$) & (R$_{\odot}$) & (M$_{\odot}$\,yr$^{-1}$) & (\kms) & (R$_{\odot}$) & (G) & (erg\,s$^{-1}$) & (erg\,s$^{-1}$)\\
  \hline
\multicolumn{11}{l}{\it Well detected objects}\\
1 & HD148937          & Of?p   & 0.39 & 1380& 5.8$\pm$0.1 & 1.8   & 4.3   & $-$5.5 & 2693 & 15  &    1.0   & 35.38    & 33.14\\ 
3 & $\theta^1$\,Ori\,C& O7Vfp  & 0.26 & 450 & 5.3$\pm$0.1 & 2.4   & 9.4   & $-$6.4 & 3225 & 9.9 &    1.1   & 34.47    & 32.99\\ 
4 & HD191612          & Of?p   & 0.32 & 2290& 5.4$\pm$0.2 & 3.7   & 57.   & $-$6.1 & 2119 & 14  &    2.5   & 34.85    & 33.35\\ 
5 & NGC1624-2         & O?p    & 0.46 & 5152& 5.1$\pm$0.2 & $>$11.& 41.   & $-$6.8 & 2890 & 9.7 & $>$20$^m$& $>$34.61 & 33.32\\ 
6 & HD47129           & O7.5III& 0.18 & 1584&5.09$\pm$0.04& $>$5.4& $<$2.2& $-$7.2 & 3567 & 10  & $>$2.8   & $>$34.00 & 32.74\\ 
7 & HD108             & Of?p   & 0.27 & 2510& 5.7$\pm$0.1 & $>$1.7& 526.  & $-$5.6 & 2022 & 19  & $>$0.50  & $>$35.00 & 32.13\\
8 & Tr16-22           & O8.5V  & 0.44 & 2290& 5.0$\pm$0.1 & $>$3.6& $<$9.9& $-$7.0 & 2742 & 9   & $>$1.5   & $>$33.94 & 32.56\\
9  & HD57682          & O9V    & 0.04 & 1300& 4.8$\pm$0.2 & 3.7   & 24.   & $-$7.1 & 2395 & 7   &    1.7   & 33.81    & 32.35\\
10 & $\zeta$\,Ori     & O9.5Ib & 0.05 & 414 & 5.6$\pm$0.1 & 1.1   & 2.1   & $-$5.9 & 1723 & 25  &    0.06  & 34.25    & $^c$  \\ 
11 & $\tau$\,sco      & B0.2V  & 0.03 & 180 & 4.5$\pm$0.1 & 1.8   & 20.   & $-$7.6 & 2176 & 5.6 &  0.20$^m$& 32.85    & 30.35\\ 
12 & NU\,Ori          & B0.5V  & 0.39 & 400 & 4.4$\pm$0.1 & 3.5   & $<$2.3& $-$8.1 & 2901 & 5.7 &    0.65  & 32.71    & 31.11\\ 
13 & HD63425          & B0.5V  & 0.06 & 1136& 4.5$\pm$0.4 & 3.1   & $<$16.& $-$7.9 & 2478 & 6.8 &    0.46  & 32.75    & 31.08\\ 
14 & HD66665          & B0.5V  & 0.012& 1500& 4.2$\pm$0.5 & 4.0   & 12.   & $-$8.2 & 2008 & 5.5 &    0.67  & 32.48    & 30.79\\
15 & $\xi^1$\,CMa     & B1III  & 0.02 & 423 & 4.6$\pm$0.1 & $>$5.3& 2.7   & $-$7.5 & 1555 & 8.6 & $>$1.5   & $>$33.17 & 31.53\\ 
17 & HD47777          & B1III  & 0.05 & 760 & 4.0$\pm$0.2 & $>$8.6& $<$4.3& $-$8.7 & 2142 & 5   & $>$2.1   & $>$32.19 & 30.51\\ 
18 & $\beta$\,Cep     & B1IV   & 0.017& 182 &4.22$\pm$0.08& 4.0   & 7.3   & $-$8.6 & 2169 & 6.5 &    0.36  & 31.99    & 30.16\\ 
20 & HD122451         & B1III  & 0    & 108 & 4.4$\pm$0.2 & $>$3.1& $<$3.2& $-$8.0 & 1552 & 8.7 & $>$0.25  & $>$32.33 & 30.34\\ 
21 & HD127381         & B1-2V  & 0.007& 127 &3.76$\pm$0.06& 7.5   & 3.8   & $-$9.7 & 2186 & 4.8 &    0.50  & 30.90    & 28.73\\
26 & HD64740          & B1.5Vp & 0.012& 350 & 4.1$\pm$0.3  & 30.  & 1.9   & $-$9.0 & 2152 & 6.3 &    16    & 32.27    & 30.43\\
28 & ALS9522          & B1.5Ve & 0.32 & 1800& 4.0$\pm$0.1 & $>$11.& $<$2.0& $-$8.0 & 989  & 6.4 & $>$4.0   & $>$32.63 & 30.59\\ 
29 & LP\,Ori          & B1.5Vp & 0.19 & 450 & 3.1$\pm$0.2 & 6.3   & $<$3.0& $-$9.0 & 758  & 2.5 &    0.91  & 31.27    & 28.60\\ 
31 & $\sigma$\,Ori\,E & B2Vp   & 0.012& 500 & 3.6$\pm$0.2 & 31.   & 2.1   & $-$9.8 & 1794 & 3.9 &   9.6$^m$& 31.29    & 29.19\\
35 & HD136504         & B2IV-V & 0.03 & 131 & 3.8$\pm$0.2 & $>$4.8& $<$5.7& $-$8.3 & 1019 & 5.3 & $>$0.60  & $>$31.99 & 29.78\\ 
39 & HD3360           & B2IV   & 0.017& 183 & 3.7$\pm$0.2 & $>$4.1& 4.4   & $-$8.4 & 942  & 5.9 & $>$0.34  & $>$31.78 & 29.34\\
42 & HD200775         & B2Ve   & 0.34 & 429 & 4.0$\pm$0.3 & 7.9   & 2.3   & $-$8.1 & 862  & 10  &    1.0   & 32.21    & 29.97\\
45 & HD182180         & B2Vn   & 0.04 & 236 & 3.0$\pm$0.1 & 41.   & 1.4   & $-$9.9 & 1058 & 3.7 &    11.   & 30.93    & 28.46\\
47 & HD142184         & B2V    & 0.08 & 130 & 2.8$\pm$0.1 & 45.   & 1.6   & $-$10.4& 1118 & 3.1 &    10.   & 30.48    & 27.92\\
63 & HD125823         & B7IIIp & 0    & 140 & 3.2$\pm$0.1 & $>$10.& 8.4   & $-$9.4 & 917  & 3.6 & $>$1.3   & $>$31.06 & 28.68\\
\hline
\multicolumn{11}{l}{\it Faint detections}\\
19 & Tr16-13          & B1V    & 0.27 & 2290& 4.0$\pm$0.1 & $>$7.7&       & $-$8.9 & 2129 & 4.9 & $>$1.4    & $>$31.92 & 30.16\\
23 & HD163472         & B1-2V  & 0.17 & 290 & 3.8$\pm$0.1 & 5.2   & 5.2   & $-$9.5 & 2466 & 4.1 & 0.40      & 31.20    & 29.09\\
27 & ALS15956         & B1.5V  & 0.17 & 5848& 4.3$\pm$0.2 & $>$9. &       & $-$8.6 & 1755 & 9.1 & $>$1.5    & $>$32.10 & 30.36\\
30 & HD37017          &B2:IV-Vp& 0.05 & 450 & 3.4$\pm$0.2 & $>$18.& 1.9   & $-$9.1 & 1102 & 3.9 & $>$6.0    & $>$31.70 & 29.57\\
61 & HD175362         & B5V    & 0.012& 275 & 3.2$\pm$0.1 & $>$59.& 3.0   & $-$9.5 & 765  & 5.8 & $>$21.$^m$& $>$31.31 & 28.50\\
\hline                                                                                                                  
\multicolumn{11}{l}{\it Undetected objects}\\                                                                           
22 & ALS3694          & B1     & 0.3  & 1750& 3.7$\pm$0.3 & $>$18.& $<$3.8& $-$8.8 & 1142 & 5.6 & $>$6   & $>$31.97 & 29.93\\
36 & HD156424         & B2V    & 0.15 & 1100& 3.7$\pm$0.4 & $>$5.2& $<$11.& $-$8.5 & 1058 & 4.8 & $>$0.65& $>$31.86 & 29.70\\
46 & HD55522          & B2IV-V &0.0004& 257 & 3.0$\pm$0.1 & $>$19.& 4.4   & $-$10.0& 1037 & 3.3 & $>$2.6 & $>$30.60 & 28.16\\
\hline
\end{tabular}
\end{table}
\end{landscape}

\setcounter{table}{0}
\begin{landscape}
\begin{table}
\footnotesize
  \caption{Continued}
  \begin{tabular}{llcccccccccccc}
  \hline
ID & Name  & sp. type & $N_{\rm H}$(ISM) & $d$ & $\log(L_{\rm BOL}/L_{\odot})$ & $R_A$ & $R_K$ & $\log(\dot M)^a$ & $v_{\infty}$ & $R_*$ & $B_p$  & $\log(L_{\rm X})$(B\&M97) & $\log(L_{\rm X})$(udd14)$^b$\\
 &  &  & (10$^{22}$\,cm$^{-2}$) & (pc) & & (R$_{\odot}$) & (R$_{\odot}$) & (M$_{\odot}$\,yr$^{-1}$) & (\kms) & (R$_{\odot}$) & (G) & (erg\,s$^{-1}$) & (erg\,s$^{-1}$)\\
  \hline
49 & HD36485          & B3Vp   & 0.03 & 524 & 3.5$\pm$0.1 & 24.   & 2.4   & $-$9.0 & 1012 & 4.5 & 10.    & 31.85    & 29.63\\
51 & HD306795         & B3V    & 0.1  & 2100& 3.2$\pm$0.3 & $>$21.& $<$3.6& $-$9.3 & 821  & 4.1 & $>$5.0 & $>$31.31 & 28.81\\
56 & HD37058          & B3VpC  & 0.012& 758 & 3.5$\pm$0.2 & $>$16.& 8.6   & $-$9.0 & 822  & 5.6 & $>$3.0 & $>$31.54 & 29.13\\
\hline
\end{tabular}

{\scriptsize Columns giving the ID, Name, sp. type, $\log(L_{\rm BOL}/L_{\odot})$, $R_A$, $R_K$, $R_*$ and $B_p$ are reproduced from Tables 1 and 6 of \citet{pet13}. The distance $d$, mass-loss rates $\log(\dot M)$, and color excesses (used to calculate the $N_{\rm H}$(ISM), see text) are the ones used in the Petit et al. paper: they were not extensively listed there and are thus reproduced here for completeness. In fact, distances $d$ and reddening for stars with luminosity determined by \citet{pet13} appear in their Table 4 and 5 while for the other objects, they were taken from the references in their Table 2. Mass-loss rates and wind velocities have been calculated using the recipes of \citet{vin00} for the chosen stellar parameters. Note that the $\dot M$ corresponds to the mass driven from an equivalent unmagnetized star, which is the parameter used in theoretical models of MCWs. Discussions on the errors of the stellar/magnetic parameters can be found in \citet[notably Sect. 3.3.3]{pet13}. Finally, the predictions $\log(L_{\rm X})$(B\&M97) and $\log(L_{\rm X})$(udd14) listed in the last two columns were calculated using the stellar/magnetic properties using the models of \citet{bab97} and \citet{udd14}, respectively (see also text for details). \\
Notes: the dubious detection of ALS8988 is not mentioned. \\
$^a$ The mass-loss rates correspond to those that the stars would have in absence of the field (which may not correspond to the actual mass-loss rate of the magnetic star, but are the parameters to be used in model calculations).\\
$^b$ These predictions have been calculated using the XADM model and a 10\% efficiency.\\
$^c$ There is no prediction for this object as the wind of the star is not confined ($\eta_*<1$).\\
$^m$ There are higher multipoles components. }
\end{table}
\end{landscape}

\clearpage
\begin{table*}
 \centering
  \caption{ONLINE MATERIAL -- List of targets, with the details of the X-ray observations. Note that, for \xmm, the quoted exposure times correspond to the shortest amongst available EPIC cameras (usually pn).}
  \label{listoftargetb}
  \begin{tabular}{llclc|llclc}
  \hline
ID & Name  & X \# & Obs. & ObsID (exp. time) & ID & Name  & X \# & Obs. & ObsID (exp. time)\\
  \hline
\multicolumn{5}{l}{\it Well detected objects}           &12 & NU\,Ori          & 4 & XMM     & 0134531601 (16ks) \\
1 & HD148937          & 1 & XMM     & 0022140101 (10ks) &12 & NU\,Ori          & 5 & XMM     & 0134531701 (21ks) \\ 
1 & HD148937          & 2 & XMM     & 0022140601 (8ks)  &12 & NU\,Ori          & 6 & Chandra & 18 (47ks)         \\
1 & HD148937          & 3 & Chandra & 10982 (99ks)      &12 & NU\,Ori          & 7 & Chandra & 1522 (38ks)       \\
3 & $\theta^1$\,Ori\,C& 1 & XMM     & 0112590301 (37ks) &12 & NU\,Ori          & 8 & Chandra & 3498 (69ks)       \\ 
4 & HD191612          & 1 & XMM     & 0300600201 (9ks)  &12 & NU\,Ori          & 9 & Chandra & 3744 (164ks)      \\ 
4 & HD191612          & 2 & XMM     & 0300600301 (12ks) &12 & NU\,Ori          &10 & Chandra & 4373 (171ks)      \\
4 & HD191612          & 3 & XMM     & 0300600401 (24ks) &12 & NU\,Ori          &11 & Chandra & 4374 (169ks)      \\
4 & HD191612          & 4 & XMM     & 0300600501 (11ks) &12 & NU\,Ori          &12 & Chandra & 4395 (100ks)      \\
4 & HD191612          & 5 & XMM     & 0500680201 (19ks) &12 & NU\,Ori          &13 & Chandra & 4396 (165ks)      \\
5 & NGC1624-2         & 1 & Chandra & 14572 (49ks)      &13 & HD63425          & 1 & XMM     & 0671990201 (18ks) \\ 
6 & HD47129           & 1 & XMM     & 0001730601 (14ks) &14 & HD66665          & 1 & XMM     & 0671990101 (25ks) \\ 
7 & HD108             & 1 & XMM     & 0109120101 (29ks) &15 & $\xi^1$\,CMa     & 1 & XMM     & 0600530101 (7ks)  \\
8 & Tr16-22           & 1 & XMM     & 0112560101 (23ks) &17 & HD47777          & 1 & XMM     & 0011420101 (31ks) \\
8 & Tr16-22           & 2 & XMM     & 0112560201 (24ks) &17 & HD47777          & 2 & XMM     & 0011420201 (34ks) \\
8 & Tr16-22           & 3 & XMM     & 0112560301 (29ks) &17 & HD47777          & 3 & Chandra & 2540 (96ks)       \\
8 & Tr16-22           & 4 & XMM     & 0112580601 (28ks) &17 & HD47777          & 4 & Chandra & 13610 (92ks)      \\
8 & Tr16-22           & 5 & XMM     & 0112580701 (8ks)  &17 & HD47777          & 5 & Chandra & 13611 (60ks)      \\
8 & Tr16-22           & 6 & XMM     & 0145740101 (7ks)  &17 & HD47777          & 6 & Chandra & 14368 (74ks)      \\
8 & Tr16-22           & 7 & XMM     & 0145740201 (7ks)  &17 & HD47777          & 7 & Chandra & 14369 (66ks)      \\
8 & Tr16-22           & 8 & XMM     & 0145740301 (7ks)  &18 & $\beta$\,Cep     & 1 & XMM     & 0300490201 (28ks) \\
8 & Tr16-22           & 9 & XMM     & 0145740401 (8ks)  &18 & $\beta$\,Cep     & 2 & XMM     & 0300490301 (29ks) \\
8 & Tr16-22           & 10& XMM     & 0145740501 (7ks)  &18 & $\beta$\,Cep     & 3 & XMM     & 0300490401 (28ks) \\
8 & Tr16-22           & 11& XMM     & 0145780101 (8ks)  &18 & $\beta$\,Cep     & 4 & XMM     & 0300490501 (28ks) \\
8 & Tr16-22           & 12& XMM     & 0160160101 (15ks) &20 & HD122451         & 1 & XMM     & 0150020101 (42ks) \\
8 & Tr16-22           & 13& XMM     & 0160160901 (31ks) &21 & HD127381         & 1 & XMM     & 0690210101 (10ks) \\
8 & Tr16-22           & 14& XMM     & 0160560101 (12ks) &26 & HD64740          & 1 & Chandra & 13625  (15ks)     \\
8 & Tr16-22           & 15& XMM     & 0160560201 (12ks) &28 & ALS9522          & 1 & Chandra & $^b$              \\
8 & Tr16-22           & 16& XMM     & 0160560301 (19ks) &29 & LP\,Ori          & 1 & Chandra & COUP (838ks)      \\
8 & Tr16-22           & 17& XMM     & 0311990101 (26ks) &31 & $\sigma$\,Ori\,E & 1 & XMM     & 0101440301 (12ks) \\
8 & Tr16-22           & 18& XMM     & 0560580101 (14ks) &31 & $\sigma$\,Ori\,E & 2 & Chandra & 3738 (91ks)       \\
8 & Tr16-22           & 19& XMM     & 0560580201 (11ks) &35 & HD136504         & 1 & XMM     & 0690210201 (5ks)  \\
8 & Tr16-22           & 20& XMM     & 0560580301 (26ks) &39 & HD3360           & 1 & XMM     & 0600530301 (14ks) \\
8 & Tr16-22           & 21& XMM     & 0560580401 (23ks) &42 & HD200775         & 1 & XMM     & 0650320101 (9ks)  \\
8 & Tr16-22           & 22& XMM     & 0650840101 (27ks) &45 & HD182180         & 1 & XMM     & 0690210401 (8ks)  \\
8 & Tr16-22           & 23& Chandra & 50 (12ks)$^a$     &47 & HD142184         & 1 & Chandra & 13624  (26ks)     \\
8 & Tr16-22           & 24& Chandra & 632 (90ks)        &63 & HD125823         & 1 & Chandra & 13618  (10ks)     \\
8 & Tr16-22           & 25& Chandra & 1249 (10ks)$^a$   &\multicolumn{5}{l}{\it Faint detections}		\\
8 & Tr16-22           & 26& Chandra & 6402 (87ks)       &19 & Tr16-13          & 1 & Chandra & CCCP              \\
8 & Tr16-22           & 27& Chandra & 11993 (44ks)      &23 & HD163472         & 1 & XMM     & 0600530201 (9ks)  \\
8 & Tr16-22           & 28& Chandra & 11994 (39ks)      &27 & ALS15956         & 1 & Chandra & CCCP              \\
9  & HD57682          & 1 & XMM     & 0650320201 (8ks)  &30 & HD37017          & 1 & XMM     & 0049560301 (14ks) \\
10 & $\zeta$\,Ori     & 1 & XMM     & 0112530101 (40ks) &61 & HD175362         & 1 & Chandra & 13619 (11ks)      \\ 
10 & $\zeta$\,Ori     & 2 & XMM     & 0657200101 (68ks) &\multicolumn{5}{l}{\it Undetected objects}		\\
10 & $\zeta$\,Ori     & 3 & XMM     & 0657200201 (33ks) &22 & ALS3694          & 1 & Chandra & 4503 (89ks)       \\
10 & $\zeta$\,Ori     & 4 & XMM     & 0657200301 (30ks) &36 & HD156424         & 1 & Chandra & 5448 (20ks)       \\
11 & $\tau$\,sco      & 1 & XMM     & 0112540101 (22ks) &46 & HD55522          & 1 & XMM     & 0690210301 (15ks) \\
12 & NU\,Ori          & 1 & XMM     & 0112590301 (38ks) &49 & HD36485          & 1 & Chandra & 639 (49ks)        \\
12 & NU\,Ori          & 2 & XMM     & 0093000101 (62ks) &51 & HD306795         & 1 & XMM     & 0201160401 (42ks) \\
12 & NU\,Ori          & 3 & XMM     & 0093000301 (17ks) &56 & HD37058          & 1 & Chandra & 2549 (49ks)       \\
\hline
\end{tabular}

{\scriptsize Note: $^a$ No correct calibration could be calculated for these two observations ; $^b$ Simultaneous fitting of 8932(30ks)+9864(24ks)+9865(17ks)+9872(9ks).}
\end{table*}

\begin{table*}
 \centering
  \caption{Luminosities (ISM absorption corrected, in the 0.5-10.0\,keV band) and \lxlb\ ratios for faint X-ray detections of magnetic stars. Upper limits of the same quantities (90\%) for non-detected objects. }
  \label{faint}
  \begin{tabular}{llccc}
  \hline
ID &   Name     & sp. type & $L_{\rm X}$ & \loglxlb \\
   &            &          & erg\,s$^{-1}$ &        \\
 \hline
\multicolumn{5}{l}{Faint detections}\\
19 &Tr16-13$^a$ &B1V           &(6.5$\pm$3.4)e29 &$-7.8\pm0.2$ \\
23 &HD\,163472  &B1/2V         &(1.2$\pm$0.2)e29 &$-8.3\pm0.1$  \\
27 &ALS15956$^a$&B1.5V         &(1.5$\pm$0.5)e31 &$-6.7\pm0.1$ \\
30 &HD\,37017   &B1.5-2.5 IV-Vp&(1.7$\pm$0.4)e29 &$-7.7\pm0.1$  \\
61 &HD\,175362  &B5V           &(7.3$\pm$3.4)e28 &$-7.9\pm0.2$ \\
\hline
\multicolumn{5}{l}{Non-detections}\\
22 &ALS3694    &B1     &$<$7.0e29 &$<-$7.4 \\
36 &HD\,156424 &B2V    &$<$6.4e29 &$<-$7.5 \\
46 &HD\,55522  &B2IV/V &$<$2.6e28 &$<-$8.1 \\ 
49 &HD\,36485  &B3Vp   &$<$5.0e28 &$<-$8.3 \\
51 &HD\,306795 &B3V    &$<$4.2e29 &$<-$7.1 \\
56 &HD\,37058  &B3VpC  &$<$1.0e29 &$<-$8.0 \\
\hline
\end{tabular}

{\scriptsize Note: $^a$ For \ch\ observations of ALS15956 and Tr16-13, we used the X-ray luminosity derived in the Carina survey \citep{naz12carina}, taking into account the different distance (and bolometric luminosities for the \loglxlb\ ratio) used here.}
\end{table*}

\begin{landscape}
\begin{table}
\scriptsize
  \caption{ONLINE MATERIAL -- Results from the spectral fits, for the models with four thermal components with temperatures fixed to 0.2, 0.6, 1.0, and 4.0\,keV. Errors equal to zero indicate fixed values, 'u' indicates an unknown value.}
  \label{4tfits}
\begin{tabular}{llcccccccccccc}
  \hline
ID & X \# & $N_{\rm H}$(add) & $norm_1$ & $norm_2$ & $norm_3$ & $norm_4$ & $\chi^2$(dof) & \multicolumn{3}{c}{$F_{\rm X}^{obs}$ (erg\,cm$^{-2}$\,s$^{-1}$)} & \multicolumn{3}{c}{$F_{\rm X}^{ISMcor}$ (erg\,cm$^{-2}$\,s$^{-1}$)} \\
 &  & (10$^{22}$\,cm$^{-2}$) & \multicolumn{4}{c}{(cm$^{-5}$)} & & 0.5--10.0\,keV & 0.5--2.0\,keV & 2.0--10.0\,keV & 0.5--10.0\,keV & 0.5--2.0\,keV & 2.0--10.0\,keV \\
  \hline
 1 &  1 & 0.36$\pm$0.00 & 1.88$\pm$0.07e-02 & 1.43$\pm$0.16e-03 & 1.65$\pm$0.09e-03 & 1.10$\pm$0.00e-03 & 1.76(314) & 3.37$\pm$0.02e-12 & 2.02$\pm$0.02e-12 & 1.35$\pm$0.01e-12 & 7.45e-12 & 6.03e-12 & 1.42e-12 \\ 
 1 &  2 & 0.36$\pm$0.00 & 1.91$\pm$0.07e-02 & 1.44$\pm$0.18e-03 & 1.67$\pm$0.09e-03 & 1.10$\pm$0.00e-03 & 1.86(303) & 3.39$\pm$0.02e-12 & 2.04$\pm$0.02e-12 & 1.35$\pm$0.01e-12 & 7.52e-12 & 6.09e-12 & 1.43e-12 \\ 
 1 &  3 & 0.36$\pm$0.00 & 1.42$\pm$0.16e-02 & 4.95$\pm$3.30e-04 & 1.74$\pm$0.16e-03 & 1.10$\pm$0.00e-03 & 1.28(248) & 2.93$\pm$0.04e-12 & 1.60$\pm$0.04e-12 & 1.33$\pm$0.01e-12 & 6.03e-12 & 4.64e-12 & 1.40e-12 \\ 
 3 &  1 & 0.53$\pm$0.02 & 7.81$\pm$1.07e-02 & 2.84$\pm$0.61e-03 & 1.00$\pm$0.05e-02 & 1.40$\pm$0.01e-02 & 7.00(326) & 2.41$\pm$0.01e-11 & 9.59$\pm$0.06e-12 & 1.45$\pm$0.01e-11 & 3.18e-11 & 1.68e-11 & 1.50e-11 \\ 
 4 &  1 & 0.36$\pm$0.00 & 2.89$\pm$0.29e-03 & 2.71$\pm$0.75e-04 & 4.00$\pm$0.44e-04 & 2.93$\pm$0.22e-04 & 1.18(167) & 8.32$\pm$0.21e-13 & 4.82$\pm$0.08e-13 & 3.50$\pm$0.20e-13 & 1.51e-12 & 1.14e-12 & 3.65e-13 \\ 
 4 &  2 & 0.36$\pm$0.00 & 2.65$\pm$0.24e-03 & 3.00$\pm$0.63e-04 & 3.21$\pm$0.37e-04 & 2.35$\pm$0.21e-04 & 1.28(166) & 7.17$\pm$0.20e-13 & 4.32$\pm$0.05e-13 & 2.85$\pm$0.20e-13 & 1.34e-12 & 1.04e-12 & 2.97e-13 \\
\hline
\end{tabular}
\end{table}
\end{landscape}
\setcounter{table}{3}
\begin{landscape}
\begin{table}
\scriptsize
  \caption{Continued}
  \begin{tabular}{llcccccccccccccccc}
  \hline
ID & X \# & $N_{\rm H}$(add) & $norm_1$ & $norm_2$ & $norm_3$ & $norm_4$ & $\chi^2$(dof) & \multicolumn{3}{c}{$F_{\rm X}^{obs}$ (erg\,cm$^{-2}$\,s$^{-1}$)} & \multicolumn{3}{c}{$F_{\rm X}^{ISMcor}$ (erg\,cm$^{-2}$\,s$^{-1}$)} \\
 &  & (10$^{22}$\,cm$^{-2}$) & \multicolumn{4}{c}{(cm$^{-5}$)} & & 0.5--10.0\,keV & 0.5--2.0\,keV & 2.0--10.0\,keV & 0.5--10.0\,keV & 0.5--2.0\,keV & 2.0--10.0\,keV \\
  \hline
12 &  1 & 0.20$\pm$0.00 & 4.09$\pm$0.63e-04 & 0.00$\pm$0.00e+00 & 5.01$\pm$0.60e-05 & 0.00$\pm$0.00e+00 & 0.82(28)  & 6.32$\pm$0.38e-14 & 5.41$\pm$0.31e-14 & 9.05$\pm$1.05e-15 & 2.33e-13 & 2.23e-13 & 9.89e-15  \\ 
12 &  2 & 0.20$\pm$0.00 & 4.78$\pm$0.51e-04 & 0.00$\pm$0.00e+00 & 4.69$\pm$0.50e-05 & 0.00$\pm$0.00e+00 & 1.21(42)  & 6.50$\pm$0.31e-14 & 5.65$\pm$0.28e-14 & 8.48$\pm$0.89e-15 & 2.54e-13 & 2.45e-13 & 9.27e-15  \\ 
12 &  3 & 0.20$\pm$0.00 & 4.09$\pm$0.97e-04 & 0.00$\pm$0.00e+00 & 5.29$\pm$0.97e-05 & 0.00$\pm$0.00e+00 & 0.98(11)  & 6.53$\pm$0.60e-14 & 5.58$\pm$0.56e-14 & 9.54$\pm$1.60e-15 & 2.37e-13 & 2.27e-13 & 1.04e-14  \\ 
12 &  4 & 0.20$\pm$0.00 & 4.24$\pm$0.52e-04 & 0.00$\pm$0.00e+00 & 4.34$\pm$0.56e-05 & 0.00$\pm$0.00e+00 & 1.81(33)  & 5.91$\pm$0.38e-14 & 5.12$\pm$0.32e-14 & 7.85$\pm$0.98e-15 & 2.26e-13 & 2.18e-13 & 8.44e-15  \\ 
12 &  5 & 0.20$\pm$0.00 & 4.20$\pm$1.04e-04 & 0.00$\pm$0.00e+00 & 6.75$\pm$1.03e-05 & 0.00$\pm$0.00e+00 & 3.03(20)  & 7.71$\pm$0.60e-14 & 6.49$\pm$0.60e-14 & 1.22$\pm$0.18e-14 & 2.64e-13 & 2.51e-13 & 1.33e-14  \\ 
12 &  6 & 0.20$\pm$0.00 & 4.89$\pm$0.95e-04 & 0.00$\pm$0.00e+00 & 3.36$\pm$0.92e-05 & 0.00$\pm$0.00e+00 & 0.93(10)  & 5.58$\pm$0.50e-14 & 4.97$\pm$0.49e-14 & 6.12$\pm$1.54e-15 & 2.39e-13 & 2.33e-13 & 6.69e-15  \\
\hline
\end{tabular}
\end{table}
\end{landscape}

\begin{landscape}
\begin{table}
 \centering
\scriptsize
  \caption{ONLINE MATERIAL -- Results from the spectral fits, for the models with free temperatures and up to two thermal components (when the normalization factors of the second component are zero, that means that the spectra were fitted with one thermal component only). Errors equal to zero indicate fixed values, 'u' indicates an unknown value.}
  \label{2tfits}
  \begin{tabular}{llcccccccccccccccc}
  \hline
ID & X \# & $N_{\rm H}$(add) & $kT_1$ & $norm_1$ & $kT_2$ & $norm_2$ & $\chi^2$(dof) & \multicolumn{3}{c}{$F_{\rm X}^{obs}$ (erg\,cm$^{-2}$\,s$^{-1}$)} & \multicolumn{3}{c}{$F_{\rm X}^{ISMcor}$ (erg\,cm$^{-2}$\,s$^{-1}$)} \\
 &  & (10$^{22}$\,cm$^{-2}$) & (keV) & (cm$^{-5}$) & (keV) & (cm$^{-5}$) & & 0.5--10.0\,keV & 0.5--2.0\,keV & 2.0--10.0\,keV & 0.5--10.0\,keV & 0.5--2.0\,keV & 2.0--10.0\,keV \\
  \hline
 1 &  1 & 0.55$\pm$0.00 & 0.20$\pm$0.00 & 7.29$\pm$0.07e-02 & 1.70$\pm$0.00 & 3.05$\pm$0.04e-03 & 2.33(315) & 3.15$\pm$0.02e-12 & 1.99$\pm$0.02e-12 & 1.16$\pm$0.03e-12 & 7.11e-12 & 5.88e-12 & 1.23e-12  \\ 
 1 &  2 & 0.55$\pm$0.00 & 0.20$\pm$0.00 & 7.36$\pm$0.08e-02 & 1.70$\pm$0.00 & 3.05$\pm$0.04e-03 & 2.63(304) & 3.16$\pm$0.03e-12 & 1.99$\pm$0.01e-12 & 1.16$\pm$0.03e-12 & 7.16e-12 & 5.92e-12 & 1.23e-12  \\ 
 1 &  3 & 0.55$\pm$0.00 & 0.20$\pm$0.00 & 5.32$\pm$0.15e-02 & 1.70$\pm$0.00 & 2.87$\pm$0.06e-03 & 1.54(249) & 2.69$\pm$0.04e-12 & 1.59$\pm$0.03e-12 & 1.10$\pm$0.03e-12 & 5.73e-12 & 4.56e-12 & 1.17e-12  \\ 
 3 &  1 & 0.66$\pm$0.01 & 0.19$\pm$0.00 & 2.70$\pm$0.24e-01 & 2.06$\pm$0.01 & 2.64$\pm$0.02e-02 & 5.95(326) & 2.22$\pm$0.01e-11 & 9.59$\pm$0.07e-12 & 1.26$\pm$0.01e-11 & 3.02e-11 & 1.72e-11 & 1.31e-11  \\ 
 4 &  1 & 0.50$\pm$0.00 & 0.24$\pm$0.00 & 6.24$\pm$0.15e-03 & 1.80$\pm$0.00 & 6.93$\pm$0.23e-04 & 1.49(169) & 7.63$\pm$0.15e-13 & 4.74$\pm$0.10e-13 & 2.89$\pm$0.12e-13 & 1.41e-12 & 1.11e-12 & 3.05e-13  \\ 
 4 &  2 & 0.50$\pm$0.00 & 0.24$\pm$0.00 & 5.90$\pm$0.13e-03 & 1.80$\pm$0.00 & 5.64$\pm$0.20e-04 & 1.57(168) & 6.61$\pm$0.12e-13 & 4.25$\pm$0.07e-13 & 2.35$\pm$0.10e-13 & 1.26e-12 & 1.01e-12 & 2.48e-13  \\
\hline
\end{tabular}
\end{table}
\end{landscape}
\setcounter{table}{4}
\begin{landscape}
\begin{table}
\scriptsize
  \caption{Continued}
  \begin{tabular}{llcccccccccccccccc}
  \hline
ID & X \# & $N_{\rm H}$(add) & $kT_1$ & $norm_1$ & $kT_2$ & $norm_2$ & $\chi^2$(dof) & \multicolumn{3}{c}{$F_{\rm X}^{obs}$ (erg\,cm$^{-2}$\,s$^{-1}$)} & \multicolumn{3}{c}{$F_{\rm X}^{ISMcor}$ (erg\,cm$^{-2}$\,s$^{-1}$)} \\
 &  & (10$^{22}$\,cm$^{-2}$) & (keV) & (cm$^{-5}$) & (keV) & (cm$^{-5}$) & & 0.5--10.0\,keV & 0.5--2.0\,keV & 2.0--10.0\,keV & 0.5--10.0\,keV & 0.5--2.0\,keV & 2.0--10.0\,keV \\
  \hline
11 &  1 & 0.00$\pm$0.01 & 0.32$\pm$0.01 & 4.48$\pm$0.13e-03 & 1.16$\pm$0.01 & 3.35$\pm$0.04e-03 & 22.4(88)  & 1.45$\pm$0.02e-11 & 1.36$\pm$0.02e-11 & 9.09$\pm$0.19e-13 & 1.61e-11 & 1.52e-11 & 9.15e-13  \\ 
12 &  1 & 0.27$\pm$0.00 & 0.19$\pm$0.00 & 7.23$\pm$1.04e-04 & 1.00$\pm$0.00 & 5.28$\pm$0.69e-05 & 0.80(28)  & 6.36$\pm$0.41e-14 & 5.42$\pm$0.38e-14 & 9.39$\pm$1.13e-15 & 2.31e-13 & 2.21e-13 & 1.03e-14  \\ 
12 &  2 & 0.27$\pm$0.00 & 0.19$\pm$0.00 & 8.33$\pm$0.84e-04 & 1.00$\pm$0.00 & 4.90$\pm$0.57e-05 & 1.21(42)  & 6.52$\pm$0.30e-14 & 5.65$\pm$0.25e-14 & 8.73$\pm$1.06e-15 & 2.51e-13 & 2.41e-13 & 9.54e-15  \\ 
12 &  3 & 0.27$\pm$0.00 & 0.19$\pm$0.00 & 7.35$\pm$1.60e-04 & 1.00$\pm$0.00 & 5.54$\pm$1.11e-05 & 0.96(11)  & 6.59$\pm$0.65e-14 & 5.60$\pm$0.50e-14 & 9.85$\pm$1.90e-15 & 2.37e-13 & 2.27e-13 & 1.08e-14  \\ 
12 &  4 & 0.27$\pm$0.00 & 0.19$\pm$0.00 & 7.38$\pm$0.86e-04 & 1.00$\pm$0.00 & 4.66$\pm$0.65e-05 & 1.73(33)  & 6.00$\pm$0.36e-14 & 5.17$\pm$0.27e-14 & 8.31$\pm$1.22e-15 & 2.25e-13 & 2.16e-13 & 8.93e-15  \\ 
12 &  5 & 0.27$\pm$0.00 & 0.19$\pm$0.00 & 7.02$\pm$1.72e-04 & 1.00$\pm$0.00 & 7.63$\pm$1.21e-05 & 2.93(20)  & 7.88$\pm$0.64e-14 & 6.53$\pm$0.54e-14 & 1.35$\pm$0.20e-14 & 2.58e-13 & 2.43e-13 & 1.48e-14  \\
\hline
\end{tabular}
\end{table}
\end{landscape}

\clearpage
\begin{landscape}
\begin{table}
 \centering
  \caption{ONLINE MATERIAL -- Main observables (average temperature, hardness ratio $=F_{\rm X}^{ISMcor} (2.0-10.0)/F_{\rm X}^{ISMcor} (0.5-2.0)$, X-ray luminosity and \loglxlb) from the spectral fits, for both models. }
  \label{observable}
  \begin{tabular}{ll|cccc|cccc}
  \hline
ID & X \# & \multicolumn{4}{|c|}{4T fits} & \multicolumn{4}{c}{2T fits}\\ 
   & & $kT_{avg}$ & HR & $ L_{\rm x}^{ISMcor,tot}$ & $\log(L_{\rm x}^{ISMcor,tot}/L_{\rm BOL})$ & $kT_{avg}$ & HR & $ L_{\rm x}^{ISMcor,tot}$ & $\log(L_{\rm x}^{ISMcor,tot}/L_{\rm BOL})$\\
   & & (keV) &  & (erg\,s$^{-1}$) & & (keV) &  & (erg\,s$^{-1}$) & \\
  \hline
 1 &  1 & 0.46$\pm$0.01 & 0.235$\pm$0.003 & 1.70e+33 & $-6.15\pm$0.003 & 0.26$\pm$0.001& 0.209$\pm$0.006 & 1.62e+33 & $-6.17\pm$0.003 \\ 
 1 &  2 & 0.46$\pm$0.01 & 0.235$\pm$0.003 & 1.71e+33 & $-6.15\pm$0.003 & 0.26$\pm$0.001& 0.208$\pm$0.005 & 1.63e+33 & $-6.17\pm$0.004 \\ 
 1 &  3 & 0.53$\pm$0.02 & 0.302$\pm$0.008 & 1.37e+33 & $-6.25\pm$0.006 & 0.28$\pm$0.001& 0.257$\pm$0.009 & 1.31e+33 & $-6.27\pm$0.006 \\ 
 3 &  1 & 0.79$\pm$0.01 & 0.893$\pm$0.008 & 7.71e+32 & $-6.01\pm$0.002 & 0.36$\pm$0.003& 0.762$\pm$0.008 & 7.32e+32 & $-6.03\pm$0.002 \\ 
 4 &  1 & 0.60$\pm$0.03 & 0.320$\pm$0.019 & 9.48e+32 & $-6.01\pm$0.01  & 0.40$\pm$0.01 & 0.275$\pm$0.013 & 8.85e+32 & $-6.04\pm$0.009 \\
\hline
\end{tabular}
\end{table}
\end{landscape}
\setcounter{table}{5}
\begin{landscape}
\begin{table}
  \caption{Continued}
  \begin{tabular}{ll|cccc|cccc}
  \hline
ID & X \# & \multicolumn{4}{|c|}{4T fits} & \multicolumn{4}{c}{2T fits}\\
   & & $kT_{avg}$ & HR & $ L_{\rm x}^{ISMcor,tot}$ & $\log(L_{\rm x}^{ISMcor,tot}/L_{\rm BOL})$ & $kT_{avg}$ & HR & $ L_{\rm x}^{ISMcor,tot}$ & $\log(L_{\rm x}^{ISMcor,tot}/L_{\rm BOL})$\\
   & & (keV) &  & (erg\,s$^{-1}$) & & (keV) &  & (erg\,s$^{-1}$) & \\
  \hline
 8 & 28 & 0.44$\pm$0.06 & 0.844$\pm$0.152 & 1.41e+32 & $-6.41\pm$0.43  & 0.69$\pm$0.05 & 0.671$\pm$0.124 & 1.36e+32 & $-6.42\pm$0.04  \\ 
 9 &  1 & 1.87$\pm$0.25 & 0.303$\pm$0.032 & 1.18e+32 & $-6.30\pm$0.02  & 1.64$\pm$0.13 & 0.249$\pm$0.012 & 1.08e+32 & $-6.34\pm$0.009 \\ 
10 &  1 & 0.32$\pm$0.003& 0.015$\pm$0.0003& 2.91e+32 & $-6.76\pm$0.002 & 0.27$\pm$0.001& 0.012$\pm$0.0002& 2.89e+32 & $-6.76\pm$0.004 \\ 
10 &  2 & 0.33$\pm$0.001& 0.016$\pm$0.0001& 2.87e+32 & $-6.77\pm$0.001 & 0.27$\pm$0.001& 0.012$\pm$0.0002& 2.87e+32 & $-6.77\pm$0.008 \\ 
10 &  3 & 0.32$\pm$0.001& 0.015$\pm$0.0001& 2.87e+32 & $-6.77\pm$0.001 & 0.27$\pm$0.001& 0.011$\pm$0.0002& 2.87e+32 & $-6.77\pm$0.004 \\
\hline
\end{tabular}
\end{table}
\end{landscape}
\setcounter{table}{5}
\begin{landscape}
\begin{table}
  \caption{Continued}
  \begin{tabular}{ll|cccc|cccc}
  \hline
ID & X \# & \multicolumn{4}{|c|}{4T fits} & \multicolumn{4}{c}{2T fits}\\
   & & $kT_{avg}$ & HR & $ L_{\rm x}^{ISMcor,tot}$ & $\log(L_{\rm x}^{ISMcor,tot}/L_{\rm BOL})$ & $kT_{avg}$ & HR & $ L_{\rm x}^{ISMcor,tot}$ & $\log(L_{\rm x}^{ISMcor,tot}/L_{\rm BOL})$\\
   & & (keV) &  & (erg\,s$^{-1}$) & & (keV) &  & (erg\,s$^{-1}$) & \\
  \hline
29 &  1 & 1.87$\pm$0.60 & 0.445$\pm$0.156 & 4.17e+28 & $-8.07\pm$0.08  & 2.60$\pm$1.91 & 0.527$\pm$0.657 & 4.51e+28 & $-8.04\pm$0.31  \\ 
31 &  1 & 1.94$\pm$0.05 & 0.308$\pm$0.008 & 9.07e+30 & $-6.21\pm$0.01  & 0.17$\pm$0.001& 0.198$\pm$0.013 & 8.77e+30 & $-6.22\pm$0.01  \\ 
31 &  2 & 2.20$\pm$0.14 & 0.372$\pm$0.030 & 7.72e+30 & $-6.28\pm$0.03  & 0.17$\pm$0.002& 0.242$\pm$0.024 & 7.27e+30 & $-6.30\pm$0.02  \\ 
35 &  1 & 3.02$\pm$0.39 & 0.694$\pm$0.050 & 1.55e+30 & $-7.16\pm$0.02  & 2.59$\pm$0.15 & 0.669$\pm$0.042 & 1.48e+30 & $-7.18\pm$0.01  \\ 
39 &  1 & 1.04$\pm$0.42 & 0.137$\pm$0.068 & 3.59e+29 & $-7.77\pm$0.04  & 0.12$\pm$0.02 & 0.020$\pm$0.004 & 2.93e+29 & $-7.86\pm$0.06  \\ 
42 &  1 & 0.26$\pm$0.01 & 0.170$\pm$0.014 & 4.25e+31 & $-5.90\pm$0.01  & 0.17$\pm$0.01 & 0.169$\pm$0.019 & 4.45e+31 & $-5.88\pm$0.03  \\ 
45 &  1 & 3.63$\pm$1.88 & 1.050$\pm$0.110 & 9.86e+29 & $-6.59\pm$0.03  & 4.38$\pm$1.26 & 1.142$\pm$0.182 & 1.09e+30 & $-6.55\pm$0.03  \\ 
47 &  1 & 1.59$\pm$0.13 & 1.611$\pm$0.135 & 1.02e+30 & $-6.35\pm$0.02  & 6.78$\pm$1.18 & 1.847$\pm$0.251 & 1.13e+30 & $-6.31\pm$0.03  \\ 
63 &  1 & 0.21$\pm$0.31 & 0.029$\pm$u     & 5.16e+28 & $-8.05\pm$u     & 0.30$\pm$0.19 & 0.011$\pm$u     & 4.88e+28 & $-8.07\pm$u     \\ 
\hline
\end{tabular}
\end{table}
\end{landscape}

\begin{table*}
 \centering
  \caption{Best-fit relations of the X-ray luminosity.}
  \label{relations}
  \begin{tabular}{l|c|c}
  \hline
Parameter & Group 1 & Group 2\\
 \hline
$\log(L_{\rm X})$ & $(0.64\pm0.06)\times \log(\dot M) + (36.7\pm0.5)$ & $(1.39\pm0.23)\times \log(\dot M) + (42.0\pm1.9)$ \\
& $(1.05\pm0.08)\times \log(L_{\rm BOL}/L_{\odot}) + (27.1\pm0.4)$ & $(1.87\pm0.26)\times \log(L_{\rm BOL}/L_{\odot}) + (22.7\pm1.0)$ \\
\hline
$\log[L_{\rm X}/L_{\rm BOL}]$ & $(0.02\pm0.05)\times \log(\dot M) + (-6.0\pm0.4)\dagger$ & $(0.81\pm0.14)\times \log(\dot M) + (-0.5\pm1.2)$ \\
& $(0.05\pm0.08)\times \log(L_{\rm BOL}/L_{\odot}) + (-6.5\pm0.4)\dagger$ & $(0.86\pm0.26)\times \log(L_{\rm BOL}/L_{\odot}) + (-10.9\pm1.0)$ \\
\hline
\end{tabular}
\\ {\scriptsize $\dagger$ Since the slopes of these relations are compatible with zero, they can be reduced to $\log[L_{\rm X}/L_{\rm BOL}]=-6.23\pm0.07$.\\
Note: We do not weight the points by the individual flux errors, notably because the errors on the other coordinate (mass-loss rate) remain uncertain. The formal errors on the fit parameters, quoted below, therefore result from scatter around the considered relation.}
\end{table*}

\begin{table*}
 \centering
  \caption{Summary of variability studies.}
  \label{variabsum}
  \begin{tabular}{llcccc}
  \hline
ID &   Name     & \multicolumn{2}{c}{\# of Obs.} & Flux var.? & Hardness var? \\
   &            & XMM & Ch.  &            &               \\
 \hline
1 & HD\,148937 & 2 & 1& N & N \\
4 & HD\,191612 & 5 & 0& Y -- 40\% & 20\% in HR, harder when brighter \\
8 & Tr16-22    & 22& 6& Y -- $\times2$ & $\times2$ in HR, harder when brighter \\
10& $\zeta$\,Ori & 4& 0& N& N\\
12& NU\,Ori    & 5 & 8& Y -- 50\% & $>\times3$ in HR, harder when brighter?\\
17& HD\,\,47777& 2 & 5& Y -- flares & N \\
18& $\beta$\,Cep & 4& 0& Y -- 10\% & N \\
31& $\sigma$\,Ori\,E & 1& 1& N & N \\
\hline
3 & $\theta^1$\,Ori\,C &&& Y -- 50\% & softer when brighter \\
\hline
\end{tabular}
\\ {\scriptsize Note: Percentage of observed flux variation correspond to $100\times(F_{max}/F_{min}-1)$, $\times2$ (resp. 3) indicates a doubling (resp. tripling) of the flux between minimum and maximum.}
\end{table*}


\begin{thebibliography}{}
\bibitem[Anders \& Grevesse(1989)]{and89} Anders, E., \& Grevesse, N.\ 1989, Geo. Chim. Acta, 53, 197 
\bibitem[Babel \& Montmerle(1997a)]{bab97} Babel, J., \& Montmerle, T.\ 1997a, A\&A, 323, 121 
\bibitem[Babel \& Montmerle(1997b)]{bab97theta} Babel, J., \& Montmerle, T.\ 1997b, ApJ, 485, L29 
\bibitem[Berghoefer et al.(1997)]{ber97} Berghoefer, T.~W., Schmitt, J.~H.~M.~M., Danner, R., \& Cassinelli, J.~P.\ 1997, A\&A, 322, 167 
\bibitem[Bohlin et al.(1978)]{boh78} Bohlin, R.~C., Savage, B.~D., \& Drake, J.~F.\ 1978, ApJ, 224, 132 
\bibitem[Bouy et al.(2009)]{bou09} Bouy, H., Hu{\'e}lamo, N., Mart{\'{\i}}n, E.~L. et al.\ 2009, A\&A, 493, 931 
\bibitem[Braithwaite(2013)]{bra13} Braithwaite, J.\ 2013, Proc. of IAUS 302, arXiv:1312.4755 
\bibitem[Cohen et al.(1997)]{coh97} Cohen, D.~H., Cassinelli, J.~P., \& Macfarlane, J.~J.\ 1997, ApJ, 487, 867 
\bibitem[Cohen et al.(2003)]{coh03} Cohen, D.~H., de Messi{\`e}res, G.~E., MacFarlane, J.~J., et al.\ 2003, ApJ, 586, 495 
\bibitem[Cohen et al.(2010)]{coh10} Cohen, D.~H., Leutenegger, M.~A., Wollman, E.~E., et al.\ 2010, MNRAS, 405, 2391 
\bibitem[Donati et al.(2006)]{don06} Donati, J.-F., Howarth, I.~D., Jardine, M.~M., et al.\ 2006, MNRAS, 370, 629 
\bibitem[Favata et al.(2009)]{fav09} Favata, F., Neiner, C., Testa, P., Hussain, G., \& Sanz-Forcada, J.\ 2009, A\&A, 495, 217 
\bibitem[Ferrario et al.(2009)]{fer09} Ferrario, L., Pringle, J.~E., Tout, C.~A., \& Wickramasinghe, D.~T.\ 2009, MNRAS, 400, L71 
\bibitem[Fossati et al.(2014)]{fos14} Fossati, L., Zwintz, K., Castro, N., et al.\ 2014, A\&A, in press, arXiv:1401.5494 
\bibitem[Gagn{\'e} et al.(1997)]{gag97} Gagn{\'e}, M., Caillault, J.-P., Stauffer, J.~R., \& Linsky, J.~L.\ 1997, ApJ, 478, L87 
\bibitem[Gagn{\'e} et al.(2005)]{gag05} Gagn{\'e}, M., Oksala, M.~E., Cohen, D.~H., et al.\ 2005, ApJ, 628, 986 (see erratum in ApJ, 634, 712)
\bibitem[Gagn{\'e} et al.(2011)]{gag11} Gagn{\'e}, M., Fehon, G., Savoy, M.~R., et al.\ 2011, ApJS, 194, 5 
\bibitem[Herv{\'e} et al.(2013)]{her13} Herv{\'e}, A., Rauw, G., \& Naz{\'e}, Y.\ 2013, A\&A, 551, A83 
\bibitem[Hubrig et al.(2011)]{hub11} Hubrig, S., et al.\ 2011, A\&A, 528, A151 
\bibitem[Ignace et al.(2010)]{ign10} Ignace, R., Oskinova, L.~M., Jardine, M., et al.\ 2010, ApJ, 721, 1412 
\bibitem[Ignace et al.(2013)]{ign13} Ignace, R., Oskinova, L.~M., \& Massa, D.\ 2013, MNRAS, 429, 516 
\bibitem[Ku et al.(1982)]{ku82} Ku, W.~H.-M., Righini-Cohen, G., \& Simon, M.\ 1982, Science, 215, 61 
\bibitem[Langer(2013)]{lan13} Langer, N.\ 2013, Proc. of IAUS 302, arXiv:1312.2373 
\bibitem[Linder et al.(2006)]{lin06} Linder, N., Rauw, G., Pollock, A.~M.~T., \& Stevens, I.~R.\ 2006, MNRAS, 370, 1623 
\bibitem[Mewe et al.(2003)]{mew03} Mewe, R., Raassen, A.~J.~J., Cassinelli, J.~P., et al.\ 2003, A\&A, 398, 203 
\bibitem[Naz{\'e} et al.(2004)]{naz04} Naz{\'e}, Y., Rauw, G., Vreux, J.-M., \& De Becker, M.\ 2004, A\&A, 417, 667 
\bibitem[Naz{\'e} et al.(2007)]{naz07} Naz{\'e}, Y., Rauw, G., Pollock, A.~M.~T., Walborn, N.~R., \& Howarth, I.~D.\ 2007, MNRAS, 375, 145 
\bibitem[Naz{\'e} et al.(2008)]{naz08} Naz{\'e}, Y., Walborn, N.~R., Rauw, G., et al.\ 2008, AJ, 135, 1946 
\bibitem[Naz{\'e}(2009)]{naz09} Naz{\'e}, Y.\ 2009, A\&A, 506, 1055 
\bibitem[Naz{\'e} et al.(2010)]{naz10} Naz{\'e}, Y., ud-Doula, A., Spano, M., et al.\ 2010, A\&A, 520, A59 
\bibitem[Naz{\'e} et al.(2011)]{naz11} Naz{\'e}, Y., Broos, P.~S., Oskinova, L., et al.\ 2011, ApJS, 194, 7 
\bibitem[Naz{\'e} et al.(2012a)]{naz12} Naz{\'e}, Y., Zhekov, S.~A., \& Walborn, N.~R.\ 2012a, ApJ, 746, 142 
\bibitem[Naz{\'e} et al.(2012b)]{naz12carina} Naz{\'e}, Y., Bagnulo, S., Petit, V., et al.\ 2012b, MNRAS, 423, 3413 
\bibitem[Naz{\'e} et al.(2012c)]{naz12zeta} Naz{\'e}, Y., Flores, C.~A., \& Rauw, G.\ 2012c, A\&A, 538, A22 
\bibitem[Naz{\'e} et al.(2013)]{naz13zeta} Naz{\'e}, Y., Oskinova, L.~M., \& Gosset, E.\ 2013, ApJ, 763, 143 
\bibitem[Naz{\'e} et al.(2014)]{naznew} Naz{\'e}, Y., Wade, G.A., \& Petit, V.\ 2014, A\&A, in press (arxiv:1408.6098)
\bibitem[Oskinova et al.(2008)]{osk08} Oskinova, L.~M., Hamann, W.-R., \& Feldmeier, A.\ 2008, Clumping in Hot-Star Winds, 203 
\bibitem[Oskinova et al.(2011)]{osk11} Oskinova, L.~M., Todt, H., Ignace, R., et al.\ 2011, MNRAS, 416, 1456 
\bibitem[Owocki et al.(2013)]{owo13} Owocki, S.~P., Sundqvist, J.~O., Cohen, D.~H., \& Gayley, K.~G.\ 2013, MNRAS, 429, 3379 
\bibitem[Petit et al.(2013)]{pet13} Petit, V., Owocki, S.~P., Wade, G.~A., et al.\ 2013, MNRAS, 429, 398 
\bibitem[Power et al.(2007)]{pow07} Power, J., Wade, G.~A., Hanes, D.~A., Aurier, M., \& Silvester, J.\ 2007, Physics of Magnetic Stars, 89 
\bibitem[Sanz-Forcada et al.(2004)]{san04} Sanz-Forcada, J., Franciosini, E., \& Pallavicini, R.\ 2004, A\&A, 421, 715 
\bibitem[Sana et al.(2006)]{san06} Sana, H., Rauw, G., Naz{\'e}, Y., Gosset, E., \& Vreux, J.-M.\ 2006, MNRAS, 372, 661 
\bibitem[Schellenberger et al.(2013)]{sch13}Schellenberger, G., Reiprich, T., \& Lovisari, L. 2013, http://web.mit.edu/iachec/meetings/2013/ Presentations/Schellenberger.pdf
\bibitem[Schulz et al.(2000)]{shu00} Schulz, N.~S., Canizares, C.~R., Huenemoerder, D., \& Lee, J.~C.\ 2000, ApJ, 545, L135 
\bibitem[Shore \& Brown(1990)]{sho90} Shore, S.~N., \& Brown, D.~N.\ 1990, ApJ, 365, 665 
\bibitem[Skinner et al.(2008)]{ski08} Skinner, S.~L., Sokal, K.~R., Cohen, D.~H., et al.\ 2008, ApJ, 683, 796 
\bibitem[Stelzer et al.(2005)]{ste05} Stelzer, B., Flaccomio, E., Montmerle, T., et al.\ 2005, ApJS, 160, 557 
\bibitem[Townsend \& Owocki(2005)]{tow05} Townsend, R.~H.~D., \& Owocki, S.~P.\ 2005, MNRAS, 357, 251 
\bibitem[Townsend et al.(2007)]{tow07} Townsend, R.~H.~D., Owocki, S.~P., \& Ud-Doula, A.\ 2007, MNRAS, 382, 139 
\bibitem[Townsend et al.(2010)]{tow10} Townsend, R.~H.~D., Oksala, M.~E., Cohen, D.~H., Owocki, S.~P., \& ud-Doula, A.\ 2010, ApJ, 714, L318 
\bibitem[Townsley et al.(2003)]{tow03} Townsley, L.~K., Feigelson, E.~D., Montmerle, T., et al.\ 2003, ApJ, 593, 874 
\bibitem[ud-Doula \& Owocki(2002)]{udd02} ud-Doula, A., \& Owocki, S.~P.\ 2002, ApJ, 576, 413 
\bibitem[ud-Doula et al.(2008)]{udd08} ud-Doula, A., Owocki, S.~P., \& Townsend, R.~H.~D.\ 2008, MNRAS, 385, 97 
\bibitem[ud-Doula et al.(2014)]{udd14} ud-Doula, A., Owocki, 
S., Townsend, R., Petit, V., \& Cohen, D.\ 2014, MNRAS, 441, 3600
\bibitem[Vink et al.(2000)]{vin00} Vink, J.~S., de Koter, A., \& Lamers, H.~J.~G.~L.~M.\ 2000, A\&A, 362, 295 
\bibitem[Wade et al.(2012a)]{wad148937} Wade, G.~A., Grunhut J., Graefener G., et al.\ 2012a, MNRAS, 419, 2459
\bibitem[Wade et al.(2013)]{wad13} Wade, G.~A., Grunhut, J., Alecian, E., et al.\ 2013, Proc. of IAUS 302, arXiv:1310.3965 
\bibitem[Wang et al.(2008)]{wan08} Wang, J., Townsley, L.~K., Feigelson, E.~D., et al.\ 2008, ApJ, 675, 464 
\bibitem[Wolff(1968)]{wol68} Wolff, S.~C.\ 1968, PASP, 80, 281 
\end{thebibliography}
\end{document}